\documentclass[12pt]{article}
\usepackage{epsfig}
\usepackage{axodraw}
\usepackage{epsfig}                            
\usepackage{graphicx}
\usepackage{rotate}
\usepackage{latexsym}
\newcommand\pubnumber{}
\newcommand\pubdate{\today}
\newcommand\hepnumber{hep-ph/0108255}

\def\csumb{Dipartimento di Fisica Teorica, Universit\`a di Torino, Italy\\
INFN, Sezione di Torino, Italy}
\def\support{\footnote{Work supported by the
European Union under contract HPRN-CT-2000-00149.}} 

\def\Title#1{\begin{center} {\Large\bf #1 } \end{center}}
\def\Author#1{\begin{center}{ \sc #1} \end{center}}
\def\Address#1{\begin{center}{ \it #1} \end{center}}

\newcommand\pubblock{\rightline{\begin{tabular}{l} \pubnumber\\
         \pubdate\\ \hepnumber \end{tabular}}}
\newenvironment{Abstract}{\begin{quotation}  }{\end{quotation}}

\makeatletter
\def\section{\@startsection{section}{0}{\z@}{5.5ex plus .5ex minus
 1.5ex}{2.3ex plus .2ex}{\large\bf}}
\def\subsection{\@startsection{subsection}{1}{\z@}{3.5ex plus .5ex minus
 1.5ex}{1.3ex plus .2ex}{\normalsize\bf}}
\def\subsubsection{\@startsection{subsubsection}{2}{\z@}{-3.5ex plus
-1ex minus  -.2ex}{2.3ex plus .2ex}{\normalsize\sl}}

\renewcommand{\@makecaption}[2]{%
   \vskip 10pt
   \setbox\@tempboxa\hbox{\small #1: #2}
   \ifdim \wd\@tempboxa >\hsize     
       \small #1: #2\par          
     \else                        
       \hbox to\hsize{\hfil\box\@tempboxa\hfil}
   \fi}

 \def\citenum#1{{\def\@cite##1##2{##1}\cite{#1}}}
\def\citea#1{\@cite{#1}{}}
 
\newcount\@tempcntc
\def\@citex[#1]#2{\if@filesw\immediate\write\@auxout{\string\citation{#2}}\fi
  \@tempcnta\z@\@tempcntb\m@ne\def\@citea{}\@cite{\@for\@citeb:=#2\do
    {\@ifundefined
       {b@\@citeb}{\@citeo\@tempcntb\m@ne\@citea\def\@citea{,}{\bf ?}\@warning
       {Citation `\@citeb' on page \thepage \space undefined}}%
    {\setbox\z@\hbox{\global\@tempcntc0\csname b@\@citeb\endcsname\relax}%
     \ifnum\@tempcntc=\z@ \@citeo\@tempcntb\m@ne
       \@citea\def\@citea{,}\hbox{\csname b@\@citeb\endcsname}%
     \else
      \advance\@tempcntb\@ne
      \ifnum\@tempcntb=\@tempcntc
      \else\advance\@tempcntb\m@ne\@citeo
      \@tempcnta\@tempcntc\@tempcntb\@tempcntc\fi\fi}}\@citeo}{#1}}
\def\@citeo{\ifnum\@tempcnta>\@tempcntb\else\@citea\def\@citea{,}%
  \ifnum\@tempcnta=\@tempcntb\the\@tempcnta\else
  {\advance\@tempcnta\@ne\ifnum\@tempcnta=\@tempcntb \else\def\@citea{--}\fi
    \advance\@tempcnta\m@ne\the\@tempcnta\@citea\the\@tempcntb}\fi\fi}
\makeatother

\newcommand{\nl}{\nonumber\\}
\newcommand{\nn}{\nonumber}
\newcommand{\ds}{\displaystyle}

\newcommand{\lpar}{\left(}                            
\newcommand{\rpar}{\right)}

\newcommand{\bq}{\begin{equation}}                    
\newcommand{\eq}{\end{equation}}
\newcommand{\bqa}{\arraycolsep 0.14em\begin{eqnarray}}
\newcommand{\eqa}{\end{eqnarray}}
\newcommand{\ba}[1]{\begin{array}{#1}}
\newcommand{\ea}{\end{array}}
\newcommand{\ben}{\begin{enumerate}}
\newcommand{\een}{\end{enumerate}}
\newcommand{\bei}{\begin{itemize}}
\newcommand{\eei}{\end{itemize}}
\newcommand{\eqn}[1]{Eq.(\ref{#1})}

\newcommand{\sect}[1]{Section~\ref{#1}}

\newcommand{\sectm}[2]{Section~\ref{#1} -- \ref{#2}}

\newcommand{\hsp}{\hspace{.5mm}}

\def\Re{\mathop{\operator@font Re}\nolimits}
\def\Im{\mathop{\operator@font Im}\nolimits}
\newcommand{\ord}[1]{{\cal O}\lpar#1\rpar}

\newcommand{\asums}[1]{\sum_{#1}}

\newcommand{\ph}{\gamma}

\newcommand{\wb}{W}

\newcommand{\zb}{Z}

\newcommand{\ff}{f}

\newcommand{\barf}{\overline f}

\newcommand{\baru}{\overline u}

\newcommand{\me}{m_e}

\newcommand{\mfs}{m^2_f}

\newcommand{\mes}{m^2_e}

\newcommand{\mlone}{m}

\newcommand{\mlones}{m^2}

\newcommand{\Mlones}{M^2}

\newcommand{\LL}{L}

\newcommand{\sman}{s}

\newcommand{\shat}{{\hat s}}
\newcommand{\that}{{\hat t}}
\newcommand{\uhat}{{\hat u}}
\newcommand{\sla}[1]{/\!\!\!#1}

\newcommand{\spro}[2]{{#1}\cdot{#2}}
\newcommand{\gadu}[1]{\gamma_{#1}}

\newcommand{\li}[2]{\mathrm{Li}_{#1}\lpar\displaystyle{#2}\rpar} 

\newcommand{\egam}[1]{\Gamma\lpar#1\rpar}               
\newcommand{\ebe}[2]{B\lpar#1,#2\rpar}                  
\newcommand{\drii}[2]{\delta_{#1#2}}                    

\newcommand{\intfx}[1]{\int_{\scriptstyle 0}^{\scriptstyle 1}\,d#1}

\newcommand{\qfs}{Q^2_f}

\newcommand{\ztwo}{\zeta(2)}

\newcommand{\bff}[4]{B_{#1}\lpar #2;#3,#4\rpar}

\newcommand{\Ddr}{{\ds\frac{1}{{\bar{\varepsilon}}}}}

\newcommand{\Ddrd}{{\bar{\varepsilon}}}
\newcommand{\ept}{\hat\varepsilon}
\newcommand{\Ddrh}{{\ds\frac{1}{\hat{\varepsilon}}}}

\newcommand{\epp}{\varepsilon'}

\newcommand{\ep}{\epsilon}

\newcommand{\tHs}{\mu}

\newcommand{\tHss}{\mu^2}
\newcommand{\Reb}{{\rm{Re}}}

\newcommand{\zvar}{z}

\newcommand{\zvars}{z^2}

\newcommand{\pmom}{p}

\newcommand{\pmoms}{p^2}

\newcommand{\kmom}{k}
\newcommand{\kmoms}{k^2}

\newcommand{\qmom}{q}
\newcommand{\qmoms}{q^2}

\newcommand{\Trmom}{Q}

\newcommand{\Trmoms}{Q^2}

\newcommand{\vect}[1]{{\vec{#1}}}

\newcommand{\upar}[1]{u}

\newcommand{\ssF}{{\scriptscriptstyle{F}}}

\newcommand{\ssI}{{\scriptscriptstyle{I}}}

\newcommand{\ssN}{{\scriptscriptstyle{N}}}

\newcommand{\ssX}{{\scriptscriptstyle{X}}}

\newcommand{\bqas}{\begin{eqnarray*}}
\newcommand{\eqas}{\end{eqnarray*}}

\def\app#1#2 {{\it Acta. Phys. Pol.} {\bf#1},#2}
\def\cpc#1#2 {{\it Computer Phys. Comm.} {\bf#1},#2}
\def\np#1#2 {{\it Nucl. Phys.} {\bf#1},#2}
\def\pl#1#2 {{\it Phys. Lett.} {\bf#1},#2}
\def\prep#1#2 {{\it Phys. Rep.} {\bf#1},#2}
\def\prev#1#2 {{\it Phys. Rev.} {\bf#1},#2}
\def\prl#1#2 {{\it Phys. Rev. Lett.} {\bf#1},#2}
\def\zp#1#2 {{\it Zeit. Phys.} {\bf#1},#2}
\def\sptp#1#2 {{\it Suppl. Prog. Theor. Phys.} {\bf#1},#2}
\def\mpl#1#2 {{\it Modern Phys. Lett.} {\bf#1},#2}
\def\jetp#1#2 {{\it Sov. Phys. JETP} {\bf#1},#2}
\def\fpj#1#2 {{\it Fortschr. Phys.} {\bf#1},#2}
\def\afp#1#2 {{\it Acta.Phys. Polon.} {\bf#1},#2}
\def\err#1#2 {{\it Erratum} {\bf#1},#2}
\def\ijmp#1#2 {{\it Int. J. Mod. Phys} {\bf#1},#2}
\def\nc#1#2 {{\it Nuovo Cimento} {\bf#1},#2}
\def\ap#1#2 {{\it Ann. Phys.} {\bf#1},#2}
\def\cmp#1#2 {{\it Comm. Math. Phys.} {\bf#1},#2}
\def\el#1#2 {{\it Europhys. Lett.} {\bf#1},#2}
\def\hpa#1#2 {{\it Helv. Phys. Acta} {\bf#1},#2}
\def\yf#1#2 {{\it Yad. Fiz.} {\bf#1},#2}
\def\nim#1#2 {{\it Nucl. Instrum. Meth.} {\bf#1},#2}
\def\spz#1#2 {{\it Sov. Pisma Zhetf} {\bf#1},#2}
\def\jetpl#1#2 {{\it JETP Lett.} {\bf#1},#2}
\def\sjnp#1#2 {{\it Sov. J. Nucl. Phys.} {\bf#1},#2}
\def\ptp#1#2 {{\it Progr. Theor. Phys. (Kyoto)} {\bf#1},#2}
\def\rmp#1#2  {{\it Rev. Mod. Phys.} {\bf#1},#2}
\def\zhetf#1#2 {{\it ZhETF} {\bf#1},#2}
\def\prs#1#2 {{\it Proc. Roy. Soc.} {\bf#1},#2}
\def\phys#1#2 {{\it Physica} {\bf#1},#2}

\newcommand{\epsi}[1]{\psi\lpar#1\rpar}               
\newcommand{\egams}[1]{\Gamma^2\lpar#1\rpar}               
\newcommand{\itpf}{\frac{1}{\lpar 2\pi\rpar^4}}

\def\bfi{\begin{figure}}
\def\efi{\end{figure}}

\begin{document}
\begin{titlepage}
\pubblock

\vfill
\def\thefootnote{\fnsymbol{footnote}}
\Title{A Practical Approach for Exponentiation\\
of QED Corrections in Arbitrary Processes} 
\vfill
\Author{Giampiero Passarino\support}
\Address{\csumb}
\vfill
\vfill
\begin{Abstract}
It is a well-known fact that, among the electroweak corrections, QED radiation
gives the largest contribution and the needed precision requires a
re-summation of the large logarithms which show up in perturbation theory.
For annihilation processes, $e^+e^- \to \barf f$, initial state radiation is
a definable, gauge-invariant, concept and one has general tools to deal with 
it; the structure function approach and also the parton-shower method. However,
when one tries to apply the algorithm to four-fermion processes that include
non-annihilation channels a problem is faced: is it still possible to
include QED corrections by making use of the standard tools?
A systematization is attempted of several, recently proposed, algorithms.
In particular, it is shown that starting from the exponentiation of soft
photons one can still derive a description of QED radiation in terms of
structure functions, i.e.\ the kernel for the hard scattering is convoluted 
with generalized structure functions where each of them is no longer function 
of one scale. Each external, charged, fermion leg brings a factor 
$x^{\scriptscriptstyle{\alpha A - 1}}$ where $\alpha$ is the fine-structure
constant, $0 \le x \le 1$ and $A$ is a function which depends on the momenta 
of the charged particles.
\end{Abstract}
\vfill
\begin{center}
\end{center}

\end{titlepage}
\def\thefootnote{\arabic{footnote}}
\setcounter{footnote}{0}
\section{Introduction}

In QED, and more generally in the standard model (SM), the most important 
terms in radiative corrections to various processes at high energy or at large 
momentum transfer are those that contain large logarithms of the type 
$\ln\lpar\Trmoms/\mlones\rpar$, where $\Trmoms$ stands for a large kinematical 
variable.
The quantity $\mlone$ is the mass of a light, charged particle, 
which emits photons, for example, the leptons or the light quarks.
The origin of these large logarithms is connected to the presence of
collinear divergences in the perturbative expansion. 
For QED in particular they arise whenever a photon is radiated in the 
direction of an incoming (outgoing) light fermion and we compute an exclusive 
observable.
Collinear divergences are present both in virtual and in real corrections.
As far as virtual corrections are concerned, they appear as a consequence of a
mass singular renormalization procedure. 
Suppose we have a Green function $F\lpar\qmom,\dots\rpar$, 
where $\qmom$ is some external momentum of a loop diagram. Then
{\em the renormalization procedure is mass singular if}
\bq
F^{\rm ren}\lpar\qmom,\dots\rpar=F\lpar \qmoms,\dots\rpar - F\lpar0,\dots\rpar,
\eq
that is, the subtraction has been performed at $\qmoms = 0$.

The collinear divergences enter $F^{\rm ren}$ through $F\lpar0,\dots\rpar$, 
in the limit $-\qmoms \gg \mlones$, where
\bq
F^{\rm ren} \sim \ln^n\frac{-\qmoms}{\mlones}\hsp,\qquad n=1,2\hsp.
\eq
When we consider, instead, the real corrections, terms like
$\Delta^{-1} = \lpar\pmom-\kmom\rpar^2+\mlones$
will arise as a consequence of a fermion of momentum $p$ and mass $\mlone$
emitting a photon of momentum $\kmom$. The singularity will arise if we
set $\mlone = 0$. With $\pmoms = \kmoms = 0$ and 
${\vect{\kmom}}\parallel{\vect{\pmom}}$
the denominator in $\Delta$ goes to zero. If we now integrate over $\kmom$,
then the
expression for the diagram in which $\Delta$ appears becomes singular.
Strictly speaking, collinear divergences are only present in theories
where massless quanta couple to each other, which is not the case for
QED, or for the SM, where the theory becomes almost massless, 
at high energies; that is, almost mass singular, due to the
presence of low mass fermions. In those cases where we need a very high
accuracy of theoretical predictions, the presence of large logarithms
calls for a re-summation of the perturbation series, in spite of the low value 
of the coupling constant. The method for this summation was actually
developed within QCD and is based on the factorization theorems, which
allow us to split the contributions of large and small distances.
As a result of this re-summation procedure one starts with the cross-section
for the so-called hard process; that is, the process with large kinematic
variables, which is subsequently convoluted with the structure function of
the initial (final) particles~\cite{sfa}.

For a given scale $\Trmom$,
the hard part of the cross-section is determined by distances of 
the order $1/\Trmom$ and is expressed in terms of the coupling constant at 
these distances, $\alpha\lpar\Trmoms\rpar$. 
This hard cross-section contains no large logarithms and is evaluated in 
perturbation theory.
The whole contribution of large distances is inserted into the 
structure functions, which obey an equation of the renormalization-group type.

The structure functions describe the radiative corrections to a given process 
in the so-called leading logarithmic approxi\-mation (LLA):
\bq
\frac{\alpha}{\pi}\,\LL = \frac{\alpha}{\pi}\,\ln\frac{\Trmoms}{\mlones}\hsp,
\eq
where $\Trmom$ represents some large scale inherent to the problem under 
consideration. Therefore, in LLA we re-sum all terms of the form
$\lpar\alpha/\pi L\rpar^n$.
Sometimes it is possible to go beyond the LLA to guarantee the correct 
evaluation of the next-to-leading terms, $\alpha/\pi\lpar\alpha/\pi L\rpar^n$. 
The latter, usually included in the so-called $K$-factor, are not universal 
but require instead a comparison with the explicit calculation of the 
cross-section up to the two-loop level. 

However, for a general process we are not always in this favorable situation.
Very often, therefore, one can find the statement that the choice of the 
appropriate scale in the structure functions is mandatory. This is a jargon 
for `implementing the correct exponentiation factor in multi-photon emission'.

The outline of the paper will be as follows: in \sect{bckg} we introduce the 
problem of a general treatment of QED radiation for arbitrary processes
which do not allow a gauge invariant definition of initial state radiation.
In \sectm{sexp}{virtual} we review the classic Yennie, Frautschi and
Suura exponentiation algorithm (hereafter YFS), completely reviewed in terms of
dimensional regularization for both virtual and real infrared terms.
In particular we give the derivation of the YFS virtual form-factor in
\sect{eval}. In \sect{tworad} we shown that the YFS real form-factor can
be exactly evaluated in $n$-dimensions without having to split a soft
part from the hard one. In \sect{twoleg} we introduce an approximation to
the exact result which is much simpler to handle in practical computations.
The extension from two emitters to an arbitrary number is discussed in
\sect{exten}. As a consequence of our approximation the result will be 
naturally expressed in terms of generalized structure functions with an
overall exponent which is discussed in \sect{sectIR}. Additional
refinements are introduced and discussed in \sect{refin}. Some general
considerations are introduced in \sect{astrat} that introduce the problems
that one encounters in going beyond the lowest order result. 
Conclusions are shown in \sect{concu}. Finally, the most relevant
integrals used in the paper are explicitly shown in appendix.

\section{Background of the problem\label{bckg}}

The great success of high precision LEP physics is intimately linked to the
possibility of splitting initial state QED radiation from the rest in a 
meaningful way and of using accurate determinations of the structure
functions for the incoming $e^+e^-$ beams, the so-called $s$-channel
structure functions~\cite{Altarelli:1989hx}.

For LEP~2 physics, although the required theoretical precision is not
comparable to what we need around the $\zb$ resonance, we are not in the
same fortunate situation~\cite{Grunewald:2000ju}. First of all, initial versus 
final state radiation
is not longer a meaningful concept. Secondly, a large fraction of processes
are not dominated by annihilation and, therefore, the standard methods
of using $s$-channel structure functions fail to reproduce the correct result.
Nevertheless, the language of structure functions is a useful one and it is
desirable to include at least the bulk of large radiative corrections.
Hence, structure functions are still applied for these processes, but
some large uncertainty remains, connected with what is usually referred to
as the problem of selecting the right scale. In a word, the choice of the 
energy scale is not a trivial issue.

For processes where some exact perturbative calculation exists the scale
inherent to structure functions can be determined by matching the two
languages but this is not possible in general. The keyword in all these
cases is, according to commonly accepted jargon, to `select a suitable
scale without knowing the exact one (two) loop calculation'. Needless
to say this attempt is utopistic, although some ingenuous strategy has
been devised in recent times. In particular we refer to some interesting
work that can be found in ref.~\cite{Kurihara:2001pe} and in
ref.~\cite{Montagna:2001ev}

More or less, all these attempts amount to start with some sort of naive
soft + virtual photon approximation, not the one commonly employed in the 
Yennie, Frautschi and Suura formalism~\cite{yfs}, but rather 
something as in the following expression:
\bqa
\frac{d\sigma_{\rm soft}(s)}{d\Omega} &=& 
\frac{d\sigma_0(s)}{d\Omega} 
\left| {\exp}\left[-\frac{\alpha}
{\pi} {\ln}\left( \frac{E}{\Delta E} \right)  
\sum_{i,j} \frac{Q_i Q_j \epsilon_i \epsilon_j}
{\beta_{ij}} {\ln}\left(\frac{1+\beta_{ij}}{1-\beta_{ij}}\right) 
\right] \right|^2,  \nl
\beta^2_{ij} &=& 1 + \lpar\frac{m_i m_j}{\spro{p_i}{p_j}}\rpar^2,
\label{attempt}
\eqa
where $m_i$ ($p_i$) are the mass(momentum) of $i$-th charged particle,
$\Delta E$ is the maximum energy of the soft photon (the boundary between 
soft- and hard-photons), $E$ is the beam energy, and $Q_i$ the electric charge 
in unit of the $e^+$ charge.
The factor $\epsilon_j$ is $-1$ for the initial particles and $+1$ for the
final particles. The indices $i,j$ run over all the charged particles 
in the initial and final states. 

We will say that any formulas as in \eqn{attempt} includes all 
{\em double-logarithms}. This terminology is not ordinarily accepted, so
we stress that it means including, prior to integration, all terms
proportional to
\bq
\ln(\Delta E/E)\,\ln(Q^2/\mfs),
\eq
where $Q^2$ denotes, generically, some large scale present in the problem.
It must be stressed that \eqn{attempt} misses to incorporate virtual
corrections and that the single-logarithmic part is omitted.
The adopted strategy is to implement a structure function 
approach~\cite{sfa} on the Born kernel for the process with the opinion that 
one is able to make an educate guess about the scale in these structure 
functions, something as in
\bqa
\sigma_{\rm tot}(s) &=& \int dx_- dy_- dx_+ dx_{u} dx_{d}   
D_{e^-}(x_-,Q^2_{\ssI -}) D_{e^-}(y_-,Q^2_{\ssF -}) \nl
{}&\times& D_{e^+}(x_+,Q^2_{\ssF +}) D_{u}(x_{u},Q^2_u) D_{d}(x_{d},
Q^2_d) \sigma_0({\hat s}), 
\eqa
where all scales, $Q^2_{\ssI -}\,$ etc are guessed. Note that, for 
definiteness, we have taken a specific example.

Another problem arises in adapting structure functions to $t$-channel
processes. A non-accelerated charged particle cannot radiate. Consider
now an incoming electron of momentum $p_{\rm in}$, an outgoing electron
of momentum $p_{\rm out}$ and the corresponding radiation of any number
of photons from both legs, before and after the hard scattering, with
total momentum $K$. Furthermore, let $Q= p_{\rm in} - p_{\rm out}$ and
consider processes dominated by the region where $Q^2 \sim 0$. The
adopted strategy would be to fold the hard scattering cross-section 
with structure functions at virtualness $Q^2$,
\bq
\beta\,\lpar 1 - x\rpar^{\beta/2 - 1}, \qquad \beta = 
\frac{2\,\alpha}{\pi}\,\lpar \ln\frac{-Q^2}{\mes} - 1\rpar,
\eq
where $x$ is the probability of finding the electron within an electron
with longitudinal momentum fraction $x$; however, the limit $Q^2 \to 0$ 
(no radiation) will not be correctly reproduced. Of course, one can switch off 
the radiation for $Q^2 \le Q^2_{\rm min}$ with $Q^2_{\rm min}$ chosen ad hoc, 
but this represents an artificial solution. Another solution is to replace the
logarithm of the structure function with a power law behavior for small
values of $Q^2$ but a continuous solution valid in any regime and not only
for $Q^2 \gg \mes$ or $Q^2 \ll \mes$ is desirable.

Furthermore, all approaches that are missing virtual photonic corrections
-- at least the universal, process independent, YFS form-factor --
simulate their effect by imposing an effective lower energy cutoff on
the photon energy, $E_{\rm c}$, and require $\ln(E_{\rm max}/E_{\rm c}) =
\ord{1}$, where $E_{\rm max}$ is the maximum photon energy.

The situation is slightly better for those $t$-channel processes where 
a perturbative calculation exists, see for instance the two-photon process
$e^+e^- \to e^+e^-\mu^+\mu^-$. Here the comparison between structure
functions and the $\ord{\alpha}$ calculation~\cite{Berends:1986if} 
allows us to select a scale $t$ once a $K$-factor is included, box diagrams 
are neglected (but luckily enough their contribution is strongly suppressed) 
and one stays away from the forward scattering region where large deviations 
are expected and seen.

Our approach is aimed to systematize this basic idea and, therefore, it must
be clearly stated that it does not represent an attempt to move beyond
the correct treatment of those terms that may lead to {\em double-logarithmic} 
enhancement. Simply, we start from the fundamental process of exponentiation 
of Yennie, Frautschi and Suura~\cite{yfs} and try to 
understand its correct interplay with the language of structure functions. 
Complete virtual corrections are not included, therefore only the universal 
YFS factor is included, collinear single-logarithms are again missing since, 
basically, only soft photons enter into the scheme.
However, the issue of `selecting the scale' turns out to be much less
arbitrary than in any previous approach. 

In particular, there is a situation where the exclusion of hard photons
represents a bad approximation to the exact result. Consider two final
state emitters, then owing to the Kinoshita, Lee and Nauenberg~\cite{KLN}
theorem (hereafter KLN) the corresponding corrections are always small and 
free of large logarithms. They can become more sizeable only for tight cuts 
on the invariant mass of the emitting pair. The typical example is in the
corrections to $e^+e^- \to \mu^+\mu^-$ where the $\ord{\alpha}$ inclusive
corrections are represented by the well-known factor $3/4(\alpha/\pi)$.
This result, however, requires adding the matrix element for single hard
bremsstrahlung and the complete $\ord{\alpha}$ virtual corrections. For a 
general process with many fermions in the final state the knowledge of this 
matrix element is usually missing and, unless tight cuts are imposed on the 
invariant masses of the final state pairs, we cannot reproduce the correct 
KLN limit.

In principle hard photons can be included for arbitrary processes in the 
so-called collinear approximation, namely hard photons are allowed only within 
a small cone surrounding each charged external fermion. In this case, however,
we are not allowed to integrate over the whole phase space of the photon
since, in this case, there is no gauge invariant leading -- sub-leading
splitting of collinear radiation and the latter, in any case, will not be
suppressed. Even more, the complete $\ord{\alpha}$ virtual corrections
are needed as well and this can only be derived on a process-by-process
basis.

YFS exponentiation has been widely used in the past~\cite{kkgang} and we have 
no pretension to be adding any substantial improvement. The only goal of this 
paper is to clarify the extraction of structure functions from the YFS program,
without having to introduce a soft-hard separation in the YFS form factor, and 
extending their validity beyond the asymptotic region where all the invariants 
of the process are much larger than all fermion masses. Therefore, the main 
difference is that we adopt a slightly different variant of the YFS-approach 
where no separation is made between the soft and the hard region; rather we 
re-formulate in modern language an old proposal by 
Chahine~\cite{Chahine:1978ai}.

\section{QED corrections}

The material in this section is well known~\cite{yfs} and we go through it 
with the main motivation of establishing notations and conventions.
As an example consider the process
\bq
e^+(p_1) + e^-(p_2) \to \barf(p_4) + f(p_3),
\label{eeff}
\eq
\subsection{Soft photon exponentiation\label{sexp}}
The cross section for the emission of an extra soft photon of momentum $k$
is
\bq
d\sigma \sim \frac{\alpha}{4\,\pi^2} \,d\sigma_0\,\int \frac{d^3k}{k_0}\,
\lpar \asums{i=1,4} \frac{\theta_ip_i}{\spro{p_i}{k}}\rpar^2,
\eq
where $d\sigma_0$ is the non-radiative differential cross-section. The
variables $\theta_i$ are defined as follows:
\bqa
\ba{lll}
{\rm in-part.}         & \quad\theta_2 & \quad -Q_e  \\
{\rm in-antipart.}     & \quad\theta_1 & \quad +Q_e  \\
{\rm out-part.}        & \quad\theta_3 & \quad +Q_f  \\
{\rm out-antipart.}    & \quad\theta_4 & \quad -Q_f  \\
\ea
\eqa
and they satisfy conservation of charge, i.e.\ $\asums{i}\,\theta_i = 0$. We 
define the usual eikonal factor,
\bq
j_{\mu}(k) = \asums{i}\,\theta_i\,\frac{p_{i\mu}}{\spro{p_i}{k}},
\label{eiko}
\eq
which satisfies current conservation,
\bq
\spro{j}{k} = \asums{i}\,\theta_i = 0.
\label{ccon}
\eq
Next we consider the generalization of process \eqn{eeff}, $e^+e^-$ 
annihilation into several fermion-antifermion pairs and an arbitrary number of
photons:
\bq
e^+(p_+)\,e^-(p_-) \to \barf_1 + f'_1 + \dots\dots + \barf_l + f'_l + n\,\ph.
\eq
In our approach intermediate vector boson are unstable particles and never
appear in the final state.
When the emitted photons are soft the corresponding amplitude is approximated,
by standard methods, and reads as follows:
\bqa
M^{\rm soft}_n &=& (i\,e)^n\,M_0\,\prod_{a=1}^{n}\,\epsilon(k_a)\cdot j(k_a),
\nl
j_{\mu}(k_a) &=& \asums{i=1,2l}\,\theta_i\,\frac{p_{i\mu}}{\spro{p_i}{k_a}},
\label{softa}
\eqa
where $\epsilon$ is the photon polarization vector, 
$\spro{\epsilon(k)}{k} = 0$, and $M_0$ is the non-radiative amplitude for the 
process. Note that, in soft approximation, only photons emitted by external
charged fermions are relevant. This fact is connected with gauge invariance
as it can be illustrated by considering $e^+e^- \to \zb\zb\gamma$. There are
two diagrams in Born approximation, direct and crossed conversions. If we 
consider only emission from the external $e^{\pm}$ lines, the Ward
identity $\spro{k}{M_{\rm ext}} = 0$ is violated. However 
$\spro{k}{M^{\rm soft}_{\rm ext}} = 0$, so that $M^{\rm soft}_{\rm ext}$ is 
gauge invariant and the full identity is restore by including $M_{\rm int}$,
i.e.\ emission from the internal electron line which, however, is neither
infrared nor collinear divergent.

The cross-section for \eqn{softa}, for an infinite number of 
emitted soft photons, is 
\bqa
\sigma &=& \asums{n=0,\infty}\,\frac{1}{n!}\,\int dPS_{\rm non-rad}\, 
dPS_{\rm rad}\,\asums{\rm spins}\,\mid M_n\mid^2\,(2\,\pi)^{-4}\, 
\delta^4\lpar p_+ + p_- - \sum_{i=1}^{2l}q_i - \sum_{j=1}^{n}k_j\rpar,  
\eqa
\bqa
dPS_{\rm non-rad} &=& (2\,\pi)^{6\,l}\prod_{i=1}^{2l}\,d^4q_i\,\delta^+\lpar 
q^2_i+m^2_i\rpar,  \nl
dPS_{\rm rad} &=& (2\,\pi)^{3\,n}\prod_{j=1}^{n}\,d^4k_j\,\delta^+\lpar 
k^2_j\rpar, \qquad \delta^+(k^2) = \theta(k_0)\,\delta(k^2),
\eqa
where we have split the total phase-space into radiation and non-radiation 
phase spaces. We easily derive
\bqa
\sigma &\sim& \sum_{n=0}^{\infty}\,\frac{1}{n!}\,\int dPS_{\rm non-rad}\, 
\asums{\rm spins}\mid M_0\mid^2\,
(2\,\pi)^{3\,n}\,\prod_{j=1}^{n}\,d^4k_j\,\delta^+(k^2_j)  \nl
{}&\times& \mid e\,j^{\mu}(k_j)\mid^2\,(2\,\pi)^{-4}\,
\delta^4\lpar p_+ + p_- - \sum_{i=1}^{2l}q_i - \sum_{j=1}^{n}k_j\rpar.
\eqa
According to the classical treatment one writes
\bqa
\delta^4\lpar K - \sum_{j=1}^{n}\,k_j\rpar &=& \int \frac{d^4x}{(2\,\pi)^4}\,
\exp\Bigl\{ i\,\spro{K}{x} - i\,\asums{j}\,\spro{k_j}{x}\Bigr\},  \nl
K &=& p_+ + p_- - \sum_{i=1}^{2l}\,q_i.
\eqa
This result can be cast into the following form:
\bq
\sigma \sim \int\,\prod_{i=1}^{2l}\,d^4q_i\,\delta^+\lpar q^2_i + m^2_i\rpar
\,(2\,\pi)^{6\,l-4}\,\asums{\rm spins}\,\mid M_0\mid^2\,E\lpar
p_+ + p_- - \asums{i}q_i\rpar,  
\eq
where the spectral function for the photon has been introduced:
\bqa
E(K) &=& \int\,\frac{d^4x}{(2\,\pi)^4}\,\sum_{n=0}^{\infty}\,\frac{E_n(x)}{n!}\,
\exp\lpar i\,\spro{K}{x}\rpar,  \nl
E_n(x) &=& \int \prod_{j=1}^n\,d^4k_j\,\delta^+(k^2_j)\,
\mid e\,j^{\mu}(k_j)\mid^2\,\exp\lpar -\,i\,\spro{k_j}{x}\rpar.
\label{spectral}
\eqa
The photons can be re-summed to all orders giving
\bqa
E(K) &=& \int\,\frac{d^4x}{(2\,\pi)^4}\,\exp\lpar i\,\spro{K}{x}\rpar\,
\sum_{n=0}^{\infty}\,\frac{1}{n!}\,\bigl[F(x)\Bigr]^n,  \nl
F(x) &=& \frac{1}{(2\,\pi)^3}\,\int d^4k\,\exp\lpar -\,i\,\spro{k}{x}\rpar\,
\delta^+(k^2)\,\mid e\,j^{\mu}(k)\mid^2.
\eqa
The Dirac delta-function, expressing four-momentum conservation, is therefore
replaced inside \eqn{spectral} by the photon spectral function.
The latter is defined through a Fourier-transform,
\bqa
E(K) &=& \frac{1}{(2\,\pi)^4}\,\int\,d^4x\,\exp\lpar i\,\spro{K}{x}\rpar\,
E(x),  \nl
E(x) &=& \exp\Bigl\{ \frac{\alpha}{2\,\pi^2}\,\int\,d^4k\,\exp\lpar -\,i\,
\spro{k}{x}\rpar\, \delta^+(k^2)\,\mid j^{\mu}(k)\mid^2\Bigr\}.
\label{defspect}
\eqa
$E(K)$ is the spectral function describing radiation and $\alpha = e^2/4\pi$
is the fine-structure constant. An important property, following from charge 
conservation is
\bq
\asums{i,j}\,\theta_i\theta_j\,\frac{\spro{p_i}{p_j}}{\spro{p_i}{k}\,
\spro{p_j}{k}} = - \asums{i<j}\,\theta_i\theta_j\,\mid 
\frac{p^{\mu}_i}{\spro{p_i}{k}} - \frac{p^{\mu}_j}{\spro{p_j}{k}}\mid^2,
\eq
and, therefore, we derive
\bq
E(x) = \exp\Bigl\{ -\,\frac{\alpha}{2\,\pi^2}\,\asums{i<j}\,\theta_i\theta_j\,
\int\,d^4k\,\exp\lpar -\,i\,\spro{k}{x}\rpar\, 
\delta^+(k^2)\,\mid\frac{p^{\mu}_i}{\spro{p_i}{k}} - 
\frac{p^{\mu}_j}{\spro{p_j}{k}}\mid^2\Bigr\}.
\eq
Note that there is no delta-function expressing four-momentum conservation
inside \eqn{spectral}, not for the full process nor for the soft limit.
Therefore, we are not allowed to use relations connected to momentum
conservation. Later in this paper we will adopt a coplanar approximation
for the exact spectral function, the difference between the two to be treated
perturbatively, which will introduce again conservation but, this time,
expressed in terms of the process where one incorporates photons emitted
along the directions of the charged fermions.
\subsection{Re-organization of the perturbative expansion\label{general}}
The exponentiation of infrared divergences is rigorous only in the limit 
where all photon momenta go to zero. The procedure adopted here will be
to define an approximation to the exact cross section where the matrix element
for the process are replaced by their soft limit. However, we do not introduce
a cutoff separating hard from soft in the exponentiation, therefore the
approximated result receives contributions from photon momenta which are large 
as well as soft. Perturbation theory is then reorganized by evaluating the
difference between the exact and the approximated contributions. The main goal 
of the present investigation will be to establish a link between our
approximation and the structure function language.
Note that, by virtue of \eqn{ccon}, our approximation satisfies $U(1)$ gauge
invariance.

Let $M_n$ denote the complete amplitude for the production of a final
state $\{q\}$ with the emission of $n$ photons,
\bq
M_n\lpar p_+,p_-;\{q\};k_1,\dots,k_n\rpar, \qquad
\rho_n = \asums{\rm spin}\,\mid M_n\mid^2.
\eq
The entire procedure amounts to construct a perturbative expansion 
which starts with an approximation that embodies the desired features
of re-summation. We introduce, as usual, the factor
\bq
J(k) = \mid e\,\spro{\epsilon(k)}{j(k)} \mid^2.
\eq
Then the complete amplitude squared can be written as
\bq
\rho_n = \beta_0\,\prod_{l=1}^{n}\,J(k_l) + \sum_{i=1}^{n}\,
\prod_{l\ne i}\,J(k_l)\beta_1(k_i) + \dots +
\beta_n\lpar k_1, \dots,k_n\rpar.
\eq
A solution for the infrared-finite residuals $\beta$ is
\bq
\beta_0 = \rho_0, \qquad \beta_1(k) = \rho_1(k) - \rho_0\,J(k), \quad
\mbox{etc.}
\label{impro}
\eq
giving a perturbative expansion for the cross-section that starts with
\bqa
\sigma &=& \frac{1}{(2\,\pi)^4}\,\int\,d^4x\,\exp\lpar i\,\spro{K}{x}\rpar\,
E(x)\,\int\,dPS_q\,\Bigl[ \beta_0  \nl
{}&+& \frac{1}{(2\,\pi)^4}\,\int\,d^4k\,\delta^+(k^2)\,\exp\lpar -\,i
\spro{k}{x}\rpar\,\beta_1(k) + \dots \bigr].
\eqa
Here $dPS_q$ is the phase-space for the final state fermions.
At this point we can ask about the validity of a result that includes only
$\beta_0$. The singular behavior of the amplitude for any radiative
process is better examined in terms of the so-called dipole formalism.
For simplicity let us consider the case of only two emitters, the 
generalization to an arbitrary number being straightforward, both outgoing.
An approximation to the true amplitude squared that incorporates the correct
singular behavior in the massless fermion limit is represented by
\bqa
\mid M\mid^2 &\sim& \mid M^{\rm non-rad}\mid^2\,g_{ij},  \nl
g_{ij} &=& \frac{z}{y\,\spro{p_i}{p_j}}\,
\Bigl[ \frac{2}{1-z(1-y)} - 1 - z\Bigr],  \nl
y &=& \frac{\spro{p_i}{k}}{\spro{p_i}{p_j}+\spro{p_i}{k}+\spro{p_j}{k}},
\qquad
z = \frac{\spro{p_i}{p_j}}{\spro{p_i}{p_j}+\spro{p_j}{k}}.
\label{softcoll}
\eqa
It is easily seen that the infrared limit corresponds to $k \to 0$ or
$y \to 0, z \to 1$ while the collinear limit to $y \to 0$, independently
of $z$. Therefore the exponentiation procedure correctly accounts
for photons that are infrared or infrared$\,\&\,$collinear but not
for photons that are hard$\,\&\,$collinear. In the coplanar approximation,
which receives contributions from photon momenta which are large 
as well as soft, we include all photons, soft or collinear that originates
from 
\bq
\frac{\spro{p_i}{p_j}}{\spro{p_i}{k}\spro{p_j}{k}},
\eq
that, however, does not reproduce the correct results of \eqn{softcoll}.
In particular $\beta_1$ is essential to correctly reproduce the KLN result.
\subsection{Inclusion of virtual corrections\label{virtual}}
Let $M_0$ be the Born amplitude for the process where $n$ photons are radiated.
Let $M_1$ be the same amplitude with one loop corrections included. Therefore
we obtain
\bqa
M &=& \exp\lpar\alpha B\rpar\,\Bigl[ M_0 + \lpar M_1 - \alpha\,B\,M_0\rpar +
\dots \Bigr],  \nl
B &=& -\,\frac{i}{8\,\pi^3}\,\asums{i<j}\,B_{ij}\,\theta_i\theta_j.
\label{virtc}
\eqa
$B_{ij}$ is the universal, i.e.\ process independent, YFS virtual factor.
Virtual corrections are operatively included through the following procedure:
\bqa
\ba{ll}
1) & \sigma \to \mid\exp\lpar \alpha\,B\rpar\mid^2\,\sigma,  \\
   &                 \\
2) & \beta_0 \to \asums{\rm spins}\,\Bigl[ \mid M_0\mid^2 + 2\,\Reb
M^*_0\,\lpar M_1 - \alpha\,B M_0\rpar\Bigr] = \beta_{00}+\beta_{01},  \\
   &                 \\
3) & \beta_1 \to \beta_1,
\ea
\eqa
where the $M_0$ in $\beta_0$ denotes the one-loop corrected amplitude with
no real photons and $\beta_1$ gives the Born amplitude with one emitted
photon. The cross-section becomes
\bqa
\sigma &=& \int\,dPS_q\,\mid \exp(\alpha\,B)\mid^2\,E\lpar p_+ + p_- -
\sum\,q\rpar\,\lpar \beta_{00} + \beta_{01}\rpar  \nl
{}&+& \frac{1}{(2\,\pi)^4}\,\int\,dPS_q\,d^4k\,\delta^+(k^2)\,d^4K\,
\mid \exp(\alpha\,B)\mid^2\,E(K)\,\frac{1}{(2\,\pi)^7}\,\int\,d^4x  \nl
{}&\times& \exp\Bigl\{i\,\lpar p_++p_- - \sum q - k - K\rpar\cdot x\Bigr\}\,
\beta_{10}(k) + \dots
\eqa
\section{Evaluation of $B_{ij}$ in dimensional regularization\label{eval}}
Consider an arbitrary process containing charged incoming and outgoing 
fermions, each with charge $Q_i$ and momentum $p_i$. The YFS 
$B_{ij}$-function in \eqn{virtc} describes infrared divergent virtual photons 
associated with the external charged lines, therefore independent of the 
internal details of the process.
Let us consider the definition of $B_{ij}$ in dimensional regularization,
i.e.\ for $n \not= 4$:
\bq
B_{ij} = \tHs^{4-n}\,\int\,\frac{d^nk}{k^2}\,{\cal B}_{ij} =
\tHs^{4-n}\,\int\,\frac{d^nk}{k^2}\,\Bigl[
{{\lpar 2\,\epsilon_i\,p_i - k\rpar_{\mu}}\over {k^2 - 2\,\epsilon_i\,
\spro{p_i}{k}}} +
{{\lpar 2\,\epsilon_j\,p_j + k\rpar_{\mu}}\over {k^2 + 2\,\epsilon_j\,
\spro{p_j}{k}}}\bigr]^2,
\eq
where $\epsilon_i = \pm 1$. 
The connection between $\epsilon_i$ and $\theta_i$ is as follows: each
particle has a charge $Q_i = 0,-1,2/3,-1/3$,
then $\theta_i = \epsilon_i\,Q_i$, or
$\epsilon_i = \pm 1$ for incoming(outgoing) fermion $i$ (outgoing(incoming)
anti-fermion $i$). Using
\bqa
k^2 - 2\,\epsilon_i\,\spro{p_i}{k} &=& \lpar k - \epsilon_i\,p_i\rpar^2 +
m^2_i \equiv (2),  \nl
k^2 + 2\,\epsilon_j\,\spro{p_j}{k} &=& \lpar k + \epsilon_j\,p_j\rpar^2 +
m^2_j \equiv (3),  
\eqa
we derive the following decomposition:
\bqa
{\cal B}_{ij} &=& \frac{1}{(2)^2}\,
\Bigl[ 4\,\lpar p^2_i - \epsilon_i\,\spro{p_i}{k}\rpar + k^2 \Bigr] + 
\frac{1}{(3)^2}\,\Bigl[ 4\,\lpar p^2_j + \epsilon_j\,\spro{p_j}{k}\rpar +
k^2 \Bigr]  \nl
{}&+& \frac{2}{(2)(3)}\,\Bigl[ 4\,\epsilon_i\epsilon_j\,\spro{p_i}{p_j} +
2\,\spro{\lpar\epsilon_i\,p_i - \epsilon_j\,p_j\rpar}{k} - k^2 \Bigr],
\eqa
which allows us to derive $B_{ij}$ in terms of standard one-loop scalar
functions~\cite{Passarino:1979jh}. We have three terms
\bq
B_{ij} = \sum_{l=1}^{3}\,B^l_{ij},
\eq
where the first one can be written as
\bq
B^1_{ij} = i\,\pi^2\,\Bigl\{ \bff{0}{p^2_i}{0}{m_i} - 2\,\frac{\partial}
{\partial m^2_i}\,\Bigl[ 2\,p^2_i\,\bff{0}{p^2_i}{0}{m_i} + p^2_i\,
\bff{1}{p^2_i}{0}{m_i}\Bigr]\Bigr\}\mid_{p^2_i= -m^2_i}.
\eq
The various terms are computed as follows~\cite{Bardin:1999ak}:
\bqa
\bff{0}{-m^2}{0}{m} &=& \Ddr - \ln\frac{m^2}{\tHss} + 2,  \nl
\frac{\partial}{\partial m^2}\,\bff{0}{p^2}{0}{m}\mid_{p^2= - m^2} &=&
- \frac{1}{2\,m^2}\,\lpar \Ddrh + \ln\frac{m^2}{\tHss}\rpar,  \nl
\frac{\partial}{\partial m^2}\,\bff{1}{p^2}{0}{m}\mid_{p^2= - m^2} &=&
\frac{1}{m^2},
\eqa
giving the first term,
\bq
B^1_{ij} = i\,\pi^2\,\Bigl[ -3\,\lpar \Ddrh + \ln\frac{m^2_i}{\tHss}\rpar + 
4\Bigr].
\eq
Similarly we derive
\bq
B^2_{ij} = i\,\pi^2\,\Bigl[ -3\,\lpar \Ddrh + \ln\frac{m^2_j}{\tHss}\rpar + 
4\Bigr].
\eq
Although not strictly necessary, we have carefully distinguished ultraviolet
($1/\Ddrd$) poles from infrared ($1/\ept$) ones,
\bq
\frac{1}{\Ddrd} = \frac{2}{\ep} - \gamma - \ln\pi, \qquad
\frac{1}{\ept} = \frac{2}{\epp} + \gamma + \ln\pi,
\eq
with regulators that satisfy $\Ddrd + \ept = 0$. The last term is the sum of 
three contributions,
\bq
B^3_{ij} = i\,\pi^2\,\Bigl\{ 8\,\epsilon_i\epsilon_j\,\spro{p_i}{p_j}\,
C_0 - 4\,P_{-ij}\cdot \Bigl[ C_{11}\,\epsilon_i p_i - C_{12}\,P_{+ij}\Bigr] 
- 2\,\bff{0}{P^2_{+ij}}{m_i}{m_j}\Bigr\},
\eq
where the $C$-functions have arguments
\bq
p_1 = - \epsilon_i p_i, \quad p_2 = P_{+ij}, \qquad m_1 = 0, m_2 = m_i, 
m_3= m_j,
\eq
and where we have introduced
\bq
P_{\pm ij} = \epsilon_i p_i \pm \epsilon_j p_j.
\eq
The function $C_0$ has a well-known representation,
\bq
C_0 = \frac{1}{2}\,\lpar F_1\,\Ddrh + F_2\rpar,  
\eq
where $F_{1,2}$ are given in terms of variables
\bq
y^{ij}_{1,2} = \frac{1}{P_{+ij}}\,\Bigl[ P^2_{+ij} + m^2_j - m^2_i \pm
\lambda^{1/2}\,\lpar - P^2_{+ij},m^2_i,m^2_j\rpar\Bigr].
\eq
Through the paper 
\bq
\lambda(x,y,z) = x^2 + y^2 + z^2 - 2\,\lpar xy + xz + yz\rpar,
\eq
represents the K\"allen's $\lambda$-function. The following result holds:
\bqa
F_1 &=& \frac{1}{P^2_{+ij}\,\lpar y^{ij}_1 - y^{ij}_2\rpar}\,\Bigl[
\ln\lpar 1 - \frac{1}{y^{ij}_2}\rpar - \ln \lpar 1 - \frac{1}{y^{ij}_1}\rpar
\Bigr],  \nl
F_2 &=& F_1\,\ln\frac{P^2_{+ij}-i\,\varepsilon}{\tHss} + 
\frac{1}{P^2_{+ij}\,\lpar y^{ij}_1 - y^{ij}_2\rpar}\,\Bigl[
f\lpar y^{ij}_1,y^{ij}_2\rpar - f \lpar y^{ij}_2,y^{ij}_1\rpar\Bigr],  \nl
f(x,y) &=& \frac{1}{2}\,\ln\lpar 1 - \frac{1}{y}\rpar\,\ln\Bigl[
y(y-1)(x-y)^2\Bigr] - \li{2}{\frac{1-y}{x-y}} + \li{2}{\frac{-y}{x-y}}.
\label{deff12}
\eqa
$\li{2}{z}$ is the standard di-logarithm.
The higher rank three-point functions can be reduced to scalar ones as
follows:
\bq
-C_{11}\,\epsilon_i p^{\mu}_i + C_{12}\,P^{\mu}_{+ij} = 
\Bigl[ {\tilde C}_{11} + C_0 \Bigr]\,\epsilon_i p^{\mu}_i + 
{\tilde C}_{12}\,\epsilon_j p^{\mu}_j.
\eq
In the above equation the ${\tilde C}$-functions have arguments
\bq
p_1 = \epsilon_i p_i, \quad p_2 = \epsilon_j p_j, \qquad
m_1 = m_i, \quad m_2 = 0, \quad m_3 = m_j.
\eq
Following standard reduction techniques we obtain
\bqa
{\tilde C}_{11} &=& -\,\frac{1}{d}\,\lpar m^2_j\,R_1 + \epsilon_i\epsilon_j\,
\spro{p_i}{p_j}\,R_2\rpar,  \nl
{\tilde C}_{12} &=& -\,\frac{1}{d}\,\lpar m^2_i\,R_2 + \epsilon_i\epsilon_j\,
\spro{p_i}{p_j}\,R_1\rpar,  \nl
R_1 &=& \frac{1}{2}\,\Bigl[ \bff{0}{P^2_{+ij}}{m_i}{m_j} - 
\bff{0}{-m^2_j}{0}{m_j} + f_1\,C_0\Bigr],  \nl
R_2 &=& \frac{1}{2}\,\Bigl[ - \bff{0}{P^2_{+ij}}{m_i}{m_j} + 
\bff{0}{-m^2_i}{0}{m_i} + f_2\,C_0\Bigr],  \nl
f_1 &=& 2\,m^2_i, \qquad f_2 = -\,2\,\epsilon_i\epsilon_j\,\spro{p_i}{p_j},
\qquad d = m^2_im^2_j - \lpar \spro{p_i}{p_j}\rpar^2,
\eqa
giving the final result for $B^3$,
\bq
B^3_{ij} = 2\,i\,\pi^2\,\Bigl[ 4\,\epsilon_i\epsilon_j\,\spro{p_i}{p_j}\,
C_0 + \bff{0}{P^2_{+ij}}{m_i}{m_j} - \sum_{l=i}^{j}\,\bff{0}{-m^2_l}{0}{m_l}
\Bigr].
\eq
If we use the explicit expressions for the scalar integrals we obtain
\bqa
C_0 &=& \frac{1}{2}\,\lpar \Ddrh + \ln\frac{P^2_{+ij}-i\,\varepsilon}
{\tHss}\rpar\,F_1 + \frac{1}{2}\,F^{\rm rest}_2,  \nl
\bff{0}{-m^2}{0}{m} &=& - \Ddrh - \ln\frac{m^2}{\tHss} + 2,  \quad
\bff{0}{P^2_{+ij}}{m_i}{m_j} = - \Ddrh - \ln\frac{m_im_j}{\tHss} + F_3,
\eqa
where $F^{\rm rest}_2$ is the part of $F_2$ not proportional to $F_1$,
see \eqn{deff12}. Furthermore, $F_3$ is the finite part of the $B_0$-function,
given by
\bq
F_3\lpar p^2;m_i,m_j\rpar = \frac{m^2_i-m^2_j}{2\,p^2}\,
\ln\frac{m^2_i}{m^2_j} + 2 -
\frac{\Lambda}{p^2}\,\ln\frac{p^2-i\,\varepsilon+m^2_i+m^2_j-\Lambda}
{2\,m_im_j},
\label{deff3}
\eq
with $\Lambda^2 = \lambda\lpar -p^2,m^2_i,m^2_j\rpar$. Using these results we 
find
\bqa
B_{ij} &=& 2\,i\pi^2\,\Bigl\{ 2\,\lpar - 1 + 
\epsilon_i\epsilon_j\,\spro{p_i}{p_j}\,
F_1\rpar\,\Ddrh - \asums{l=ij}\,\ln\frac{m^2_i}{\tHss}  \nl
{}&+& 2\,\epsilon_i\epsilon_j\,\spro{p_i}{p_j}\,\Bigl[ F_1\,
\ln\frac{P^2_{+ij}-i\,\varepsilon}{\tHss} + F^{\rm rest}_2\Bigr] + F_3 \Bigr\},
\eqa
showing that the virtual infrared pole originates from a $C_0$ function.
$F_1$ and $F^{\rm rest}_2$ are defined in \eqn{deff12}, $F_3$ in \eqn{deff3}.
Let $Q_{ij} = P_{+ij}$, we will consider two limiting cases.
\subsection{The case of large invariant}
If $m_i = m_j = m$ the results simplify into
\bqa
F_1 &=& \frac{2}{P^2_{+ij}\beta_{ij}}\,\ln\frac{\beta_{ij}+1}{\beta_{ij}-1},
\nl
F^{\rm rest}_2 &=& \frac{1}{P^2_{+ij}\beta_{ij}}\,\Bigl[
\ln\frac{\beta_{ij}+1}{\beta_{ij}-1}\,\ln\frac{m^2\beta^2_{ij}}{P^2_{+ij}} -
2\,\sum_{l=\pm 1}\,\li{2}{\frac{\beta_{ij}+l}{2\,\beta_{ij}}}
\Bigr],
\eqa
with $\beta^2_{ij} = 1 +4\,m^2/P^2_{+ij}$. Furthermore
\bq
F_3 = 2 - \beta_{ij}\,\ln\frac{\beta_{ij}+1}{\beta_{ij}-1}.
\eq
In the limit $\mid Q_{ij}\mid \gg m^2$, where $Q_{ij} = P_{+ij}$, we obtain
\bqa
F_1 &\sim& \frac{2}{Q^2_{ij}}\,\ln\frac{Q^2_{ij}-i\epsilon}{m^2},  \nl
F^{\rm rest}_2 &=& -\,\frac{1}{Q^2_{ij}}\,\Bigl[
\ln^2\frac{Q^2_{ij}-i\epsilon}{m^2} + \frac{\pi^2}{3}\Bigr],  \nl
F_3 &\sim& -\,\ln\frac{Q^2_{ij}-i\epsilon}{m^2} + 2.
\label{asympt}
\eqa
\subsection{The case of small invariant}
Consider now the opposite limit where $\mid Q^2_{ij}\mid \ll m^2_i,m^2_j$.
We easily derive
\bq
F_1 \sim \frac{1}{m^2_j-m^2_i}\,\ln\frac{m^2_j}{m^2_i}, \qquad
F_2 \sim \frac{1}{2}\,\frac{1}{m^2_j-m^2_i}\,\Bigl[
\ln^2\frac{m^2_j}{\tHss} - \ln^2\frac{m^2_i}{\tHss}\Bigr].
\eq
Note that, for $m_i = m_j = m$ this further simplifies into
\bq
F_1 \sim \frac{1}{m^2}, \qquad F_2 \sim \frac{1}{m^2}\,\ln\frac{m^2}{\tHss}.
\eq
Furthermore the two-point function is
\bq
\bff{0}{0}{m_i}{m_j} = - \Ddrh - 1 - \frac{1}{m^2_j-m^2_i}\,\Bigl[
m^2_j\,\ln\frac{m^2_j}{\tHss} - m^2_i\,\ln\frac{m^2_i}{\tHss}\Bigr],
\eq
which, for equal masses, gives
\bq
\bff{0}{0}{m}{m} = - \Ddrh - \ln\frac{m^2}{\tHss}.
\eq
Collecting the various terms and introducing $r= m^2_j/m^2_i$ we obtain
\bqa
B_{ij} &=& i\,\pi^2\,\Bigl\{\Bigl[ \frac{r+1}{r-1}\,\ln r - 2\Bigr]\,
\Ddrh + \frac{r+1}{r-1}\,\Bigl[ \ln^2\frac{m^2_j}{\tHss} - 
\ln^2\frac{m^2_i}{\tHss}\Bigr]  \nl
{}&+& \frac{3-r}{r-1}\,\ln\frac{m^2_i}{\tHss} + \frac{1-3\,r}{r-1}\,
\ln\frac{m^2_j}{\tHss} + 2 \Bigr\},
\eqa
which, for equal masses gives
\bq
B_{ij}\lpar Q^2,m_i=m_j=m\rpar \to 0, \qquad \mbox{for}\quad Q^2 \to 0.
\eq
\section{The two-particle radiation factor\label{tworad}}
The photon spectral function is defined in \eqn{defspect}. Remarkably
enough the exponent in $E(x)$ can be computed exactly~\cite{Chahine:1978ai}. 
We will show it by computing $E(x)$ for the case of only two emitters.
This underlying ingredient will be referred to as the two-particle radiation 
factor, ${\cal R}_{ij}$. In dimensional regularization it reads as follows:
\bq
{\cal R}_{ij} = \tHs^{4-n}\,\int\,d^nk\,\exp\lpar -\,i\,\spro{k}{x}\rpar\,
\frac{\delta^+(k^2)}{\spro{p_i}{k}\spro{p_j}{k}},
\eq
where $\tHs$ is the arbitrary unit of mass and
\bq
p^2_i = - m^2_i, \qquad q^2_{ij} = \lpar p_i+p_j\rpar^2.
\eq
Furthermore $\delta^+$ selects positive energies,
\bq
\delta^+(k^2) = \theta(k_0)\,\delta(k^2).
\eq
In order to evaluate ${\cal R}_{ij}$ we introduce a parameter $\rho_{ij}$ 
defined by the following relations~\cite{'tHooft:1979xw}:
\bq
p = \rho_{ij}\,p_i, \qquad q = p_j, \qquad \lpar p - q\rpar^2 = 0.
\label{defrho}
\eq
Hence we have a solution
\bq
\rho^{\pm}_{ij} = \frac{1}{2\,m^2_i}\,\Bigl[ -q^2_{ij} - m^2_i - m^2_j \pm
\lambda^{1/2}\lpar -q^2_{ij},m^2_i,m^2_j\rpar\Bigr].
\eq
If $q^2_{ij} < 0$ we select $\rho^+$, while $\rho^-$ is chosen when 
$q^2_{ij} > 0$. Next we introduce a Feynman parameter $u$ and define
\bq
P = q + \lpar p - q\rpar\,u.
\eq
With our choice for $\rho$ it follows that $P^2 < 0$ and we will compute
${\cal R}$ in the frame where 
\bq
{\bf P} = 0, \qquad P_0 = M.
\eq
Therefore we obtain
\bqa
{\cal R} &=& \rho\,\intfx{u}\,J,  \nl
J &=& \frac{\pi^{n/2-1}\tHs^{4-n}}{M^2\,\egam{n/2-1}}\,\int_0^{\infty}\,
dk\,k^{n-5}\,\int_{-1}^{+1}\,dy\,\lpar 1 - y^2\rpar^{n/2-2}\,
\exp\{ -\,i\,k\,\lpar yr - x_0\rpar\},
\nl
\eqa
where $r = |{\bf x}|$. If we write
\bq
{\cal R} = \rho\,\frac{\pi^{n/2-1}\tHs^{4-n}}{\egam{n/2-1}}\,\int_0^1\,
\frac{du}{M^2}\,H,
\eq
then the function $H$ becomes
\bqa
H &=& \int_0^{\infty}\,dk\,k^{n-5}\,\int_{-1}^{+1}\,dy\,
\exp\{ -\,i\,k\lpar yr-x_0\rpar\}\,\lpar 1 - y^2\rpar^{n/2-2}  \nl
{}&=& \lpar\frac{2}{r}\rpar^{n/2-3/2}\,\pi^{1/2}\,\egam{n/2-1}\,
\int_0^{\infty}\,dk\,\exp\lpar \,i\,kx_0\rpar\,k^{n/2-7/2}\,J_{n/2-3/2}(kr),
\eqa
where $J_{\nu}$ is a Bessel-function. As a consequence we arrive at the 
following result:
\bqa
{\cal R} &=& 2^{n/2-3/2}\,\rho\,\pi^{n/2-1/2}\,
\lpar \tHs r\rpar^{4-n}\,\intfx{u}\,\frac{K}{M^2},  \nl
K &=& K_{\rm r} + i\,\epsilon(x_0)\,K_{\rm i} =
\int_0^{\infty}\,dk\,\exp\lpar i\,\frac{x_0}{r}k\rpar\,k^{n/2-7/2}\,
J_{n/2-3/2}(k).
\label{defK}
\eqa
Real and imaginary parts of ${\cal R}$ will be computed separately.
\subsection{The real part}
We start by computing the real part of $K$, \eqn{defK},
\bq
K_{\rm r} = \int_0^{\infty}\,dk\,\cos\lpar \xi k\rpar\,k^{n/2-7/2}\,
J_{n/2-3/2}(k), \quad \xi = \frac{x_0}{r}, \quad 0 < \xi < \infty.
\eq
This function will be considered separately in the two regions $0 < \xi < 1$
and $1 < \xi < \infty$. For the former we find
\bq
0 < \xi < 1, \qquad K_{\rm r} = 2^{n/2-7/2}\,\frac{\egam{n/2-2}}{\egam{3/2}}\,
{}_2F_1\,\lpar \frac{n-4}{2},-\frac{1}{2};\frac{1}{2};\xi^2\rpar,
\eq
where ${}_2F_1$ is the standard hypergeometric function~\cite{bath}. Let 
$n = 4 + \epp$, with $\epp \ge 0$, then we can use the Laurent's expansion of 
the $\Gamma$-function
\bq
\egam{\frac{\epp}{2}} = \frac{2}{\epp} - \gamma + \ord{\epp},  
\eq
and one of the transformation properties of the hypergeometric function
to arrive at the following form:
\bq
{}_2F_1\lpar \frac{\epp}{2},-\frac{1}{2};\frac{1}{2};\xi^2\rpar =
\lpar 1 - \xi^2\rpar^{1-\epp/2}\,{}_2F_1\lpar \frac{1-\epp}{2},1;
\frac{1}{2};\xi^2\rpar.
\eq
Since we are in the region $0 \le \xi^2 \le 1$ the following relation holds:
\bq
{}_2F_1\lpar \frac{1-\epp}{2},1;\frac{1}{2};\xi^2\rpar = 
\frac{\egam{1/2}}{\egam{1/2-\epp/2}}
\, \sum_{l=0}^{\infty}\,\frac{
\egam{l+1/2-\epp/2}\egam{l+1}}{\egam{l+1/2}}\,\frac{\xi^{2l}}{l!}.
\eq
The function $\egam{\epp/2}$ shows an infrared pole and, therefore, we must 
expand the hypergeometric function in powers of $\epp$. The whole procedure is 
cumbersome and, essentially, requires to obtain derivative of ${}_2F_1$ with 
respect to the parameters. Using the following expansion for the Euler 
gamma-function
\bq
\egam{a+\lambda\,\epp} = \egam{a}\,\Bigl[ 1 + \lambda\,\psi(a)\,\epp +
\ord{\epp^2}\Bigr],
\eq
where it appears the Euler $\psi$-function, we obtain
\bq
{}_2F_1\lpar \frac{1-\epp}{2},1;\frac{1}{2};\xi^2\rpar = 
\sum_{l=0}^{\infty}\,\frac{\egam{2+l}}{l!}\,\Bigl\{ 1  
+ \frac{\epp}{2}\,\Bigl[ \epsi{\frac{1}{2}} - \epsi{\frac{1}{2}+l}
\Bigr]\Bigr\}\,\xi^{2l} + \ord{\epp^2}.
\eq
As a consequence we are to consider the following expansion
\bqa
{}_2F_1\lpar \frac{1-\epp}{2},1;\frac{1}{2};\xi^2\rpar &=& 
\lpar 1 - \xi^2\rpar^{-1} + \frac{\epp}{2}\,E_1(\xi^2) + \ord{\epp^2},  \nl
E_1(\xi^2) &=& 
\sum_{l=0}^{\infty}\,\Bigl[ \epsi{\frac{1}{2}} - \epsi{\frac{1}{2}+l}\Bigr]
\,\xi^{2l}.
\eqa
The series in $E_1$ can be re-summed as follows. First we write
\bq
\epsi{l+\frac{1}{2}} - \epsi{\frac{1}{2}} = \intfx{x}\,x^{-1/2}\,
\frac{1-x^l}{1-x},
\eq
and successively we obtain
\bqa
\sum_{l=0}^{\infty}\,\Bigl[\epsi{l+\frac{1}{2}} - \epsi{\frac{1}{2}}\Bigr] \,
z^l &=& \sum_{l=0}^{\infty}\,\intfx{x}\,x^{-1/2}\,
\frac{1-x^l}{1-x}\,z^l,  \nl
{}&=& \frac{z}{1-z}\,\intfx{x}\,x^{-1/2}\,\lpar 1 - zx\rpar^{-1}
\eqa
The re-summation gives a rather simple result, 
\bq
E_1(\xi) = -\,\frac{\xi}{1-\xi^2}\,\,\ln\frac{1+\xi}{1-\xi}.
\eq
For $K_{\rm r}$ in this region we find
\bq
K_{\rm r} = \lpar 2\,\pi\rpar^{-1/2}\,\Bigl[ \frac{2}{\epp} - \gamma +
\ln 2 - \ln\lpar 1 - \xi^2\rpar - \xi\,\ln\frac{1+\xi}{1-\xi}\Bigr].
\eq
The result for ${\cal R}_{\rm r}$ in the region $0 \le \xi \le 1$ is
\bq
{\cal R}_{\rm r} = \pi\rho\,\int_0^1 \frac{du}{M^2}\,
\Bigl[ \frac{2}{\epp} - \gamma + \ln\pi + 2\,\ln 2 -
\ln \tHss x^2 - \xi\,\ln\frac{1+\xi}{1-\xi}\Bigr].
\eq
Now we turn to the complementary region $1 \le \xi \le \infty$, where
\bq
K_{\rm r} = 2^{3/2-n/2}\xi^{4-n}\,\cos\lpar \frac{n-4}{2}\,\pi\rpar\,
\frac{\egam{n-4}}{\egam{n-1/2}}\,{}_2F_1\lpar \frac{n}{2}-2,
\frac{n-3}{2};\frac{n-1}{2};\xi^{-2}\rpar,
\eq
and where we will use
\bq
{}_2F_1\lpar \frac{\epp}{2},\frac{1+\epp}{2};\frac{3+\epp}{2};\xi^{-2}\rpar =
\lpar 1 - \frac{1}{\xi^2}\rpar^{1-\epp/2}\,{}_2F_1\lpar \frac{3}{2},1;
\frac{3+\epp}{2};\xi^{-2}\rpar.
\eq
Since we are in the region where $0 \le \xi^{-2} \le 1$ the following
result holds:
\bqa
{}_2F_1\lpar \frac{3}{2},1;\frac{3+\epp}{2};\xi^{-2}\rpar
&=& \lpar 1 - \frac{1}{\xi^2}\rpar^{-1} - \frac{\epp}{2}\,E_2(\xi) +
\ord{\epp^2},  \nl
E_2(\xi) &=& \sum_{l=0}^{\infty}\,\Bigl[ \epsi{l+\frac{3}{2}} - 
\epsi{\frac{3}{2}}\Bigr]\,\xi^{-2l}.
\eqa
The series for $E_2$ can be re-summed by using the following identity:
\bq
\epsi{\frac{3}{2}+l} - \epsi{\frac{3}{2}} = \intfx{x}\, x^{1/2}\,
\frac{1-x^l}{1-x},  
\eq
and, consequently, we get
\bqa
\sum_{l=0}^{\infty}\,\Bigl[ \epsi{l+\frac{3}{2}} -
\epsi{\frac{3}{2}}\Bigr]\,z^l &=& 
\sum_{l=0}^{\infty}\,\intfx{x}\,x^{1/2}\,\frac{1-x^l}{1-x}\,z^l  \nl
{}&=& \frac{z}{1-z}\,\intfx{x}\,\frac{x^{1/2}}{1-zx}.
\eqa
We obtain the following result for $E_2$:
\bq
E_2(\xi) = \lpar 1 - \frac{1}{\xi^2}\rpar^{-1}\,\Bigl[
\xi\,\ln\frac{\xi+1}{\xi-1} - 2\Bigr],
\eq
and
\bq
K_{\rm r} = \lpar\frac{2}{\pi}\rpar^{1/2}\,\Bigl[ \frac{1}{\epp} - 
\frac{1}{2}\,\gamma + \frac{1}{2}\,\ln 2 - \frac{1}{2}\,\ln\lpar 1 - 
\frac{1}{\xi^2}\rpar - \frac{1}{2}\,\xi\,\ln\frac{\xi+1}{\xi-1}\Bigr].
\eq
As far as the real part is concerned we have
\bqa
0 \le \xi \le 1  &{}& {\cal R}_{\rm r} = \pi\rho\,\int_0^1 \frac{du}{M^2}\,
\Bigl[ \frac{2}{\epp} - \gamma + \ln\pi + 2\,\ln 2 -
\ln \tHss x^2 - \xi\,\ln\frac{1+\xi}{1-\xi}\Bigr],  \nl
1 \le \xi \le \infty  &{}& {\cal R}_{\rm r} = \pi\rho\,\int_0^1\,
\frac{du}{M^2}\,\Bigl[ \frac{2}{\epp} - \gamma + \ln\pi + 2\,\ln 2 -
\ln(- \tHss x^2) - \xi\,\ln\frac{\xi+1}{\xi-1}\Bigr].
\nl
\eqa
\subsection{The imaginary part}
The imaginary part of the two-particle radiator is given in terms of
\bq
K_{\rm i} = \int_0^{\infty}\,dk\,\sin\lpar \xi k\rpar\,k^{n/2-7/2}\,
J_{n/2-3/2}(k).
\eq
This can again be written in terms of hypergeometric functions,
\bqa
0 \le \xi \le 1  &{}& K_{\rm i} = 
2^{n/2-5/2}\,\xi\,\egam{\frac{n-3}{2}}\,{}_2F_1\lpar \frac{n-3}{2},0;
\frac{3}{2};\xi^2\rpar,  \nl
1 \le \xi \le \infty  &{}& K_{\rm i} = 2^{3/2-n/2}\,\xi^{4-n}\,
\frac{\egam{n-4}}{\egam{n/2-1/2}}\,\sin\frac{(n-4)\,\pi}{2}  \nl
{}&\times& {}_2F_1\lpar \frac{n-3}{2},\frac{n-4}{2};\frac{n-1}{2};\xi^{-2}
\rpar.
\eqa
We use the following expansion,
\bq
\sin\frac{(n-4)\,\pi}{2}\,\egam{n-4} = \frac{\pi}{2}\,\egam{n-3} + \ord{n-4},
\eq
to show that, as expected, the imaginary part has no infrared poles and,
therefore, we may set $n = 4$. The result is
\bqa
0 \le \xi \le 1  &{}& K_{\rm i} = 2^{-1/2}\,\pi^{1/2}\,\xi\,
{}_2F_1\lpar \frac{1}{2},0;\frac{3}{2};\xi^2\rpar,  \nl
1 \le \xi \le \infty  &{}& K_{\rm i} = 2^{-1/2}\,\pi^{1/2}
\,{}_2F_1\lpar \frac{1}{2},0;\frac{3}{2};\xi^{-2}
\rpar.
\eqa
It is straightforward to derive that
\bq
{}_2F_1\lpar \frac{1}{2},0;\frac{3}{2};\xi^2\rpar = 1.
\eq
The final result for the imaginary part reads as follows:
\bqa       
0 \le \xi \le 1  &{}& {\cal R}_{\rm i} = \rho\,\pi^2\,\int_0^1\,
\frac{du}{M^2}\,\xi,  \nl
1 \le \xi \le \infty  &{}& {\cal R}_{\rm i} = \rho\,\pi^2\,\int_0^1\,
\frac{du}{M^2}.
\eqa
In order to cast the final result into a more compact form we introduce an 
infinitesimal quantity $\delta$ such that
\bq
x_0 \to x_0 + i\,\delta, \qquad \delta \to 0_+,
\eq
It follows that $x^2 \to r^2 - x^2_0 - i\,x_0\delta$ and
\[
\ln(x^2) \to \ln\lpar x^2 - i\,x_0\,\delta\rpar = 
\left\{
\ba{ll}
\ln(x^2) & \mbox{for} \quad x^2 > 0 \\
\ln(-x^2) - i\,\pi\,\epsilon(x_0) & \mbox{for} \quad x^2 < 0 
\ea
\right.
\]
\subsection{The complete result}
Collecting the results for ${\cal R} = {\cal R}_{\rm r} + i\,\epsilon(x_0)\,
{\cal R}_{\rm i}$, we obtain for $0 \le \xi \le 1$
\bqa
{\cal R}_{\rm r} &=& \rho\pi\,\int_0^1\,\frac{du}{M^2}\, 
\Bigl[ \frac{1}{\ept} + 2\,\lpar \ln 2 - \gamma\rpar - \ln \tHss x^2 - 
\xi\,\ln\frac{1+\xi}{1-\xi}\Bigr],
\nl
{\cal R}_{\rm i} &=& \rho\pi^2\,\int_0^1\,\frac{du}{M^2}\,\xi, 
\eqa
and for $1 \le \xi \le \infty$,
\bqa
{\cal R}_{\rm r} &=& \rho\pi\,\int_0^1\,\frac{du}{M^2}\, 
\Bigl[ \frac{1}{\ept} + 2\,\lpar \ln 2 - \gamma\rpar - \ln(- \tHss x^2) - 
\xi\,\ln\frac{\xi+1}{\xi-1}\Bigr],
\nl
{\cal R}_{\rm i} &=& \rho\pi^2\,\int_0^1\,\frac{du}{M^2}.
\eqa
Here we have introduced
\bq
{\ds\frac{2}{\varepsilon'}} = \Ddrh - \gamma - \ln\pi. 
\eq
The total result is
\bq
{\cal R} = {\cal R}_{\rm r} + i\,\epsilon(x_0)\,{\cal R}_{\rm i}.
\eq
For $1 \le \xi \le \infty$ it follows $x^2 \le 0$ and, therefore 
\bq
\ln(-x^2) - i\,\pi\epsilon(x_0) \to \ln x^2,
\eq
while, for $0 \le \xi \le 1$ we can replace
\bq
\ln\frac{1+\xi}{1-\xi} + i\,\pi\epsilon(x_0) \to \ln\frac{\xi+1}{\xi-1}.
\eq
Therefore the function ${\cal R}$ is defined on the whole $\xi$-axis
by the following expression:
\bqa
{\cal R} &=& \rho\,\pi\,\int_0^1\,\frac{du}{M^2}\,\Bigl[ \frac{1}{\ept} + 
2\,\lpar \ln 2 - \gamma \rpar - \ln \tHss x^2 - \xi\,
\ln\frac{\xi+1}{\xi-1}\Bigr]  
\nl
{} &=& \rho\,\pi\,\int_0^1\,\frac{du}{M^2}\,\Bigl[ \Delta_{\rm IR}  - 
\ln \tHss x^2 - \xi\,\ln\frac{\xi+1}{\xi-1}\Bigr],  
\nl
\Delta_{\rm IR} &=& \frac{1}{\ept} + 2\,\lpar \ln 2 - \gamma\rpar,  \nl
\xi &=& \frac{x_0}{r}, \qquad x_0 = x_0 + i\,\delta, \quad \delta \to 0_+.
\eqa
To get the final form of our result we must express all quantities in 
covariant form. Therefore we have
\bq
M^2 = - P^2, \quad x_0 = - \frac{\spro{P}{x}}{M}, \quad
\xi^2 = \frac{\lpar\spro{P}{x}\rpar^2}{\lpar\spro{P}{x}\rpar^2 + P^2x^2}.
\label{cov}
\eq
\section{Radiation from two legs\label{twoleg}}
Consider the case when the photon can be emitted by two external charged 
fermions only. The corresponding radiator is
\bq
{\cal R}\lpar p_i,p_j\rpar = - \frac{\theta_i\theta_j}{2\,\pi^2}\,\Bigl[
p^2_i\,{\cal R}_{ii} + p^2_j\,{\cal R}_{jj} - 2\,\spro{p_i}{p_j}\,
{\cal R}_{ij}\Bigr],
\eq
and we need the photon spectral function,
\bq
E_{ij}(x) \equiv E\lpar p_i,p_j;x\rpar = 
\frac{1}{(2\,\pi)^4}\,\int\,d^4K\,\exp\Bigl\{ -\,i\,\spro{K}{x} + 
\alpha\,{\cal R}\lpar p_i,p_j\rpar\Bigr\}.
\label{spect}
\eq
We rewrite the radiator as
\bqa
{\cal R}_{ij} &=& \rho_{ij}\,\pi\,\lpar \Delta_{\rm IR} - \ln \tHss x^2\rpar\,
\int_0^1\,\frac{du}{M^2} + \rho_{ij}\,\pi\,r_{ij},  \nl
{}&=& \rho_{ij}\,\pi\,\lpar \Delta_{\rm IR} - \ln \tHss x^2\rpar\,
\frac{1}{q^2-p^2}\,\ln\frac{p^2}{q^2} + \rho_{ij}\,\pi\,r_{ij},  \nl
{}&=& -\,\rho_{ij}\,\pi\,\ln\lpar e^{-\Delta_{\rm IR}}\,\tHss x^2\rpar\,
\frac{1}{q^2-p^2}\,\ln\frac{p^2}{q^2} + \rho_{ij}\,\pi\,r_{ij}.
\eqa
In principle there is no problem in evaluating the finite part of the radiator
explicitly, with a result that contains several di-logarithms. However,
we have to exponentiate it as in \eqn{spect} and, successively, we must
compute the Fourier transform of the result: we will not be able to proceed 
any further with the complete expression.
Using \eqn{cov} we get
\bq
P^2 = q^2 + \lpar p^2 - q^2\rpar\,u,  \qquad
\spro{P}{x} = \spro{q}{x} + \spro{(p-q)}{x}\,u,
\eq
and $\xi$ becomes
\bqa
\xi^2 &=& \frac{\lpar a + b\,u\rpar^2}{A\,u^2 + 2\,B\,u + C},  \qquad
a = \spro{q}{x}, \qquad b = \spro{(p-q)}{x},  \nl
C &=& \lpar\spro{q}{x}\rpar^2, \quad B = \spro{q}{x}\,\spro{(p-q)}{x} +
\frac{1}{2}\,\lpar p^2 - q^2\rpar\,x^2, \quad A = b^2.
\eqa
Therefore, one integral is immediate
\bq
\int_0^1\,\frac{du}{M^2} = \frac{1}{p^2-q^2}\,\ln\frac{p^2}{q^2},
\eq
while the remaining one starts with
\bq
\intfx{u}\,\frac{\xi}{M^2}\,\ln\frac{\xi+1}{\xi-1}  
= -\,\intfx{u}\,\frac{a+bu}{\lpar c+du\rpar\,U}\,
\ln\frac{a+bu+U}{a+bu-U},
\eq
where $U^2 = Au^2+2 Bu+C$ and, moreover, $c = q^2, d = p^2-q^2$. 
If needed the last integral can be computed through the substitution
\bq
t = \frac{u}{U-\sqrt{C}}.
\eq
\subsection{Coplanar approximation}
The Fourier transform of \eqn{spect} cannot be computed in closed form, 
therefore we change strategy and introduce an approximated formulas which
is much simpler to handle in practical computations~\cite{Chahine:1978ai}. 
This approximation is the coplanar one, where the effective photon momentum is 
constrained to lie in the plane formed by $p_i$ and $p_j$, so that the 
spectral function turns out to be proportional to $\delta^2(K_{\perp})$, where 
$K_{\perp}$ is the transverse component of $K$. This coplanar approximation 
reads
\bq
{\cal R}^{\rm c}\lpar p_i,p_j\rpar = -\,\frac{\theta_i\theta_j}{\pi}\,
\Bigl\{ \ln\lpar -\,e^{-\Delta^{\rm c}_{\rm IR}}\,\frac{\tHss\,
\spro{p_i}{x}\spro{p_j}{x}}{s_{ij}}\rpar\,\Bigl[ 1 + \rho_{ij}\,
\frac{\spro{p_i}{p_j}}{p^2_j-\rho^2_{ij}\,p^2_i}\,
\ln\frac{\rho^2_{ij}\,p^2_i}{p^2_j}\Bigr] +
\frac{1}{2} + \frac{\pi^2}{6}\Bigr\},
\label{defcop}
\eq
where we have used the fact that for $i = j$ $\rho_{ii} = 1, M^2 = m^2_i$.
Several new quantities have been introduced,
\bq
\Delta^{\rm c}_{\rm IR} = \frac{1}{\ept} - 2\,\gamma + \frac{3}{2}, \quad
s_{ij} = \Bigl[ 1 + \frac{\mid Q^2_{ij}\mid}{m_im_j}\Bigr]^{1/2}\,m_im_j, 
\quad
Q^2_{ij} = \lpar \epsilon_i p_i + \epsilon_j p_j\rpar^2.
\label{defdelta}
\eq
Note that $s_{ij}$ satisfies the following asymptotic behavior.
\bqa
s_{ij} &\sim& \Bigl[ \mid Q^2_{ij}\mid m_im_j\Bigr]^{1/2}, \qquad
\mid Q^2_{ij}\mid \gg m^2_i,m^2_j,  \nl
s_{ij} &\sim& m_im_j, \qquad 
\mid Q^2_{ij}\mid \ll m^2_i,m^2_j.
\label{sasym}
\eqa
The same result is rewritten as
\bq
{\cal R}^{\rm c}\lpar p_i,p_j\rpar = - A_{ij}\,\ln\lpar
-\,e^{-\Delta^{\rm c}_{\rm IR}}\, \frac{\tHss\,\spro{p_i}{x}{p_j}{x}}
{s_{ij}}\rpar + \delta_{ij},
\eq
with a function $A_{ij}$ defined by
\bq
A_{ij} = \frac{\theta_i\theta_j}{\pi}\,\Bigl[ 1 - \rho_{ij}\,
\frac{\spro{p_i}{p_j}}{m^2_j-\rho^2_{ij}\,m^2_i}\,
\ln\frac{\rho^2_{ij}\,m^2_i}{m^2_j}\Bigr],  \quad
\delta_{ij} = -\,\frac{\theta_i\theta_j}{\pi}\,\lpar \frac{1}{2} +
\frac{\pi^2}{6}\rpar.
\eq
There is some element of ambiguity in the definition of the coplanar
factor of \eqn{defcop}, which disappears when the difference (exact - 
coplanar) is properly included. It remains, however, when the result
is expressed solely in terms of the coplanar approximation. This is connected
to the fact that collinear logarithms, e.g.\ $\ln(Q^2/m^2)$, are not fully 
accounted in the exponentiation and only double-logarithms, of the form
$\ln(\Delta E/E)\ln(Q^2/m^2)$, are properly included.

In other words, what we can do is as follows:

\begin{enumerate}

\item to exponentiate according to the YFS recipe and we do that, in
principle, by including the full eikonal factor, no soft limit;

\item to compensate with respect to the complete answer by including
infrared safe residuals, $\beta_1$ or more, see \eqn{impro};

\item to translate the bulk of exponentiation into structure functions
(see \eqn{strf} below) including the rest into a remainder 
(see \eqn{sfflux} below) which could be handled numerically.

\end{enumerate}

From this point of view it really does not matter which approximation we 
start with, the only problem being that we, in general, do not control
$\beta_1$ and neglect remainders. This is why the coplanar approximation is 
aimed to be as accurate as possible.

The main characteristics that an approximation to the exact spectral function
has to satisfy are: a) the possibility of extracting the typical form of
the solution of the evolution equations for fermion(anti-fermion) distributions
in the soft limit and b) the correct exponentiation of the leading
logarithms. Furthermore, it should respect the correct scaling behavior and
return no radiation for $p_i = p_j$. The particular choice made in 
\eqn{defdelta} will be commented in \sect{sectIR}.

If we now write
\bq
A_{ij} = \theta_i\theta_j\,{\cal A}_{ij},
\eq
it is easily seen that ${\cal A}_{ij}$ is non-positive $\forall i,j$. Indeed
we can immediately derive that
\bqa
{\cal A}_{ij} &\propto& -\,{\bf k}^2\,\int\,d\Omega_k\,{\cal A}^{\mu}_{ij}
{\cal A}_{\mu ij},  \nl
{\cal A}_{\mu ij} &=& \lpar\frac{p_{i\mu}}{\spro{p_i}{k}} - 
\frac{p_{j\mu}}{\spro{p_j}{k}}\rpar,
\label{spacel}
\eqa
where the integration is over the angular variables of the photon.
From $\spro{{\cal A}_{ij}}{k} = 0$ and $k^2 = 0$ it follows that 
$\spro{{\cal A}_{ij}}{{\cal A}_{ij}} \ge 0$. 
In coplanar approximation we have
\bq
E^{\rm c}\lpar p_i,p_j;K\rpar =
\frac{1}{(2\,\pi)^4}\,\int\,d^4x\,\exp\Bigl\{ i\,\spro{K}{x} +
\alpha\,{\cal R}^{\rm c}(p_i,p_j)\Bigr\}.
\eq
The total radiator will be the sum of its coplanar approximation and a 
remainder,
\bq
{\cal R}\lpar p_i,p_j\rpar = {\cal R}^{\rm c}\lpar p_i,p_j\rpar +
{\cal R}^{\rm rest}\lpar p_i,p_j\rpar.
\eq
In this way we obtain the spectral function $E(K)$ as
\bqa
E_{ij}(K) &=& E\lpar p_i,p_j;K\rpar =
\frac{1}{(2\,\pi)^4}\,\int\,d^4x\,\exp\Bigl\{ i\,\spro{K}{x} +
\alpha\,{\cal R}^{\rm c}(p_i,p_j) + 
\alpha\,{\cal R}^{\rm rest}(p_i,p_j)\Bigr\}  \nl
{}&=& \int\,d^4K'\,\Phi\lpar K-K'\rpar\,
\frac{1}{(2\,\pi)^4}\,\int\,d^4x\,\exp\Bigl\{ i\,\spro{K'}{x} +
\alpha\,{\cal R}^{\rm c}\Bigr\},  \nl
\Phi(K) &=& \frac{1}{(2\,\pi)^4}\,\int\,d^4y\,\exp\Bigl\{ i\,\spro{K}{y} +
\alpha\,{\cal R}^{\rm rest}\Bigr\}.
\label{sfflux}
\eqa
The exact spectral function is now written as the convolution of some
flux $\Phi$ with a kernel integral that we may cast in an appropriate
form
\bqa
{\cal H}(K) &=&  \frac{1}{(2\,\pi)^4}\,\int\,d^4x\,\exp\Bigl\{ i\,
\spro{K}{x} + \alpha\,{\cal R}^{\rm c}\Bigr\}  \nl
{}&=& \frac{1}{(2\,\pi)^4}\,\int\,d^4x\,\exp\lpar i\,\spro{K}{x} +
\alpha\,\delta_{ij}\rpar\,
\Bigl[ -\,e^{-\Delta^{\rm c}_{\rm IR}}\,\frac{\tHss\,\spro{p_i}{x}\spro{p_j}{x}}
{s_{ij}}\Bigr]^{-\alpha\,A_{ij}}.
\eqa
The flux-function can be expanded in powers of $\alpha$, giving
\bq
\Phi(K) = \delta^4(K) + \frac{\alpha}{(2\,\pi)^4}\,\int\,d^4y\,
\exp\lpar i\,\spro{K}{y}\rpar\,{\cal R}^{\rm rest} + \ord{\alpha^2}.
\eq
For the kernel ${\cal H}$ we use the fact that $x_0$ is defined with a small
imaginary part, or
\bq
i\,\spro{p}{x} \to i\,\spro{p}{x} + \delta.
\eq
Futhermore, the relation
\bq
\lpar i\,\spro{p}{x}\rpar^{-s} = \frac{1}{\egam{s}}\,\int_0^{\infty}\,
d\sigma\,\sigma^{s-1}\,\exp\lpar -\,i\,\spro{p}{x}\,\sigma\rpar,
\label{gir}
\eq
is used to obtain
\bqa
{\cal H}(K) &=& \frac{1}{(2\,\pi)^4}\,\int\,d^4x\,
\exp\lpar i\,\spro{K}{x} + \alpha\,\delta_{ij}\rpar\,
\Bigl[-e^{-\,\Delta^{\rm c}_{\rm IR}}\,\frac{\tHss\,\spro{p_i}{x}\spro{p_j}{x}}
{s_{ij}}\Bigr]^{-\alpha\,A_{ij}} \nl
{}&=& \Bigl[ e^{-\Delta^{\rm c}_{\rm IR}}\,\frac{\tHss}{s_{ij}}\Bigr]^
{-\alpha A_{ij}}\,
\frac{e^{\alpha\delta_{ij}}}{\egams{\alpha A_{ij}}}\,
\int_0^{\infty}\,d\sigma d\sigma'\,
\lpar \sigma\sigma'\rpar^{\alpha A_{ij} - 1}\,
\delta^4\lpar \sigma p_i + \sigma' p_j - K\rpar.
\label{strf}
\eqa
\eqn{strf} exhibits the typical form of a structure function, i.e.\
\bq
\frac{\alpha A_{ij}}{\egam{\alpha A_{ij} + 1}}\,\sigma^{\alpha A_{ij} -1}.
\eq
Note that the quantity $\Delta^{\rm c}_{\rm IR}$ inside \eqn{strf} exhibits 
the infrared pole.
\eqn{gir} is valid for $\Reb s > 0$ and, therefore, \eqn{strf} is valid
only if $\Reb A_{ij} > 0$, which is not always the case due to the
$\theta_i\theta_j$ factor. We will examine in the next section the 
generalization to positive exponents.

The key relation, therefore, is \eqn{gir} which expresses the factor
$i\,\spro{p}{x}$ through the Mellin transform of an exponential that,
in turn, re-establish energy-momentum conservation explicitly. For
alternative uses of Mellin type representations see the first of 
ref.~\cite{kkgang}.

To summarize our findings the coplanar approximation to the exact
photon spectral function is given by
\bqa
E^{\rm c}\lpar p_i,p_j;K\rpar &=& 
F_{\rm IR}\,\int\,d^4K'\int_0^{\infty}\,d\sigma_id\sigma_j\,
\Phi(K')\,\Bigl[ \frac{\alpha A_{ij}}{\egam{\alpha A_{ij} + 1}}\Bigr]^2\,
\lpar \sigma_i\sigma_j\rpar^{\alpha A_{ij} - 1}  \nl
{}&\times& \delta^4\lpar \sigma_i p_i + \sigma_j p_j - K -K'\rpar,  \nl
F_{\rm IR} &=& \Bigl[ e^{-\Delta^{\rm c}_{\rm IR}}\,\frac{\tHss}
{s_{ij}}\Bigr]^{-\alpha A_{ij}}\,e^{\alpha\delta_{ij}}.
\label{summar}
\eqa
There is an important property that any spectral function must satisfy,
corresponding to the fact that a non-accelerated charged particle cannot 
radiate. From \eqn{eiko} and from \eqn{spacel} it follows that both the
exact and the coplanar spectral function satisfy
\bq
E^{\rm ex,c}\lpar p,p;K\rpar \propto \delta^4(K),
\eq
since both $j^{\mu}$ and ${\cal A}^{\mu}_{ii}$ are zero.

A convenient inclusion of the flux-function $\Phi$ towards the exact
spectral function is achievable with the
help of \eqn{summar}: let us assume that both $p_i$ and $p_j$ 
denote outgoing momenta, then according to a well-know strategy which is 
adopted in the dipole formalism~\cite{Catani:1997vz}, we define transformed 
momenta ${\tilde p}_j, {\tilde p_j}$. Let us introduce the following 
quantities:
\bq
\lambda_{ij} = \lambda\lpar -K^2,\sigma^2_i m^2_i,\sigma^2_j m^2_j\rpar,
\quad
\lambda_j = \lambda\lpar -\lpar \sigma_i p_i + K'\rpar^2, -K^2,
\sigma^2_j m^2_j\rpar,
\eq
then the transformed momenta are
\bqa
{\tilde p}_{i\mu} &=& \sigma_i p_{i\mu} + K'_{\mu} + \lpar \sigma_j p_{j\mu} -
{\tilde p}_{j\mu}\rpar,  \nl
{\tilde p}_{j\mu} &=& \lpar \frac{\lambda_{ij}}{\lambda_j}\rpar^{1/2}\,
\sigma_j \lpar p_{j\mu} - \frac{\spro{K}{p_j}}{K^2}\,K_{\mu}\rpar 
+ \frac{1}{2}\,\frac{K^2-\sigma^2_j m^2_j + \sigma^2_i m^2_i}{K^2}\,
K_{\mu}.
\eqa
By construction they satisfy
\bq
\sigma_i p_i + \sigma_j p_j + K' = {\tilde p}_i + {\tilde p}_j = K
\eq
and
\bq
{\tilde p}^2_i = - \sigma^2_i m^2_i, \qquad
{\tilde p}^2_j = - \sigma^2_j m^2_j.
\eq
Moreover, if the phase space element is written as
\bq
d\phi\lpar p_i,p_j;K';K\rpar = \prod_{l=1,j}\,d^4p_l\,
\delta^+(p^2_l+m^2_l)\,\delta^4\lpar \sigma_i p_i + \sigma_j p_j - K -K'\rpar,
\eq
after the transformation we get the following decomposition,
\bq
d\phi\lpar p_i,p_j;K';K\rpar = d\phi\lpar {\tilde p}_i,{\tilde p}_j;0;K\rpar 
\otimes [dK'],
\eq
which implies
\bqa
E\lpar p_i,p_j;K\rpar &=& 
F_{\rm IR}\,\int_0^{\infty}\,d\sigma_id\sigma_j\,
\Bigl[ \frac{\alpha A_{ij}}{\egam{\alpha A_{ij} + 1}}\Bigr]^2\,
(\sigma_i\sigma_j)^{\alpha A_{ij} - 1}\,\Phi_{\rm int}\,
\delta^4\lpar {\tilde p_i} + {\tilde p}_j - K\rpar,  \nl
\Phi_{\rm int} &=& \int\,[dK']\,\Phi(K').
\label{exactsf}
\eqa
The explicit form of the integration measure $[dK']$ depends on the nature
of $p_i, p_j$, i.e.\ incoming and/or outgoing and the successive integration
represents the difficult part of the procedure. We will not investigate it
any further.

Note that the coplanar approximation satisfies the scaling property
\bq
E^{\rm c}\lpar p_i,p_j;\lambda K\rpar = \lambda^{2\,\alpha A_{ij} - 4}\,
E^{\rm c}\lpar p_i,p_j;K\rpar,
\eq
and it is symmetric, i.e.\ $E^{\rm c}(p_i,p_j;K) = E^{\rm c}(p_j,p_i;k)$.

An additional comment concerns the correct interpretation of the parameters 
$\sigma,\sigma'$. In the standard approach the structure function represents
the probability of finding an electron (or any fermion) within an
electron (or any fermion) with a longitudinal momentum fraction $\sigma$.
In the standard approach, therefore, the emitted photons, typically
representing initial state radiation in $e^+e^-$-annihilation are strictly
collinear. There is a subtle point here~\cite{Passarino:1998ts}: consider 
$e^+e^-$ annihilation processes, dominated or not by $s$-channel diagrams.
Let $p$ be the four-momentum of the incoming electron in the laboratory
system,
\bq
p = \frac{1}{2}\sqrt{s}\,\lpar 0,0,\beta,1\rpar, \qquad \beta^2= 1 - 4\,
\frac{\mes}{s}.
\label{defbeta}
\eq
The electron, before interacting, emits soft and collinear photons. Let $k=
k_1+k_2+\dots$ be the total four-momentum of the radiated photons. Thus
\bq
k= \frac{1}{2}\sqrt{s}\,(1-x)\,\lpar 0,0,1,1\rpar,
\eq
so that $k^2= 0$, as requested by collinear, massless, photons.
Usually one can work with the massless approximation for the electron taking
part in the hard scattering, thus an on-shell (massless) electron can
emit a bunch of massless, collinear, photons and remain on its (massless)
mass shell. But the electron mass cannot be neglected in particular cases
and, after radiation, the electron finds itself in a virtual state having 
four-momentum
\bq
{\hat p} = p - k = \frac{1}{2}\sqrt{s}\,\lpar 0,0,\beta-1+x,x\rpar,
\eq
with $x$ being the fraction of energy remaining after radiation. As a 
consequence, the electron is put off its mass shell,
\bq
{\hat p}^2 = - \mes + \frac{1}{2}(1-\beta)(1-x)\,s \sim -x\,\mes \quad
\mbox{for} \quad \me \to 0.
\eq
When considering the whole process we introduce $p_{\pm}$ for the incoming 
$e^{\pm}$ in the laboratory system. Once radiation has been emitted
the momenta will be denoted by ${\hat p}_{\pm}$ with
\bq
{\hat p}_{\pm} = \frac{1}{2}\sqrt{s}\,\lpar 0,0,\mp(\beta-1+x_{\pm}),
x_{\pm}\rpar.
\eq
The total four-momentum becomes
\bq
{\hat P} = {\hat p}_{+} + {\hat p}_{-} = \frac{1}{2}\sqrt{s}\,\lpar 0,0,
x_- - x_+, x_- + x_+\rpar,
\eq
with a corresponding invariant mass ${\hat P}^2 = - x_+x_-\,s = \shat$.

In our approach things are different since $K$ inside \eqn{exactsf} is,
by no means, restricted by the condition $K^2 = 0$. In other words $K$,
instead of being collinear, is coplanar and we have no problem in
dealing with relations like ${\hat p} = (1-x)\,p$ and $p^2= -\mes$
simultaneously. Needless to say coplanar includes collinear.
\section{Extension to $n$ emitters\label{exten}}
The result of the previous sector should now be generalized to an arbitrary
number of emitters.
With $n$ emitters we have $n(n-1)/2$ pairs and $n(n-1)$ $\sigma$-variables.
The corresponding spectral function becomes
\bq
E(K) = \frac{1}{(2\,\pi)^4}\,\int\,d^4x\,\exp\lpar i\,\spro{K}{x}\rpar\,
\prod_{l+1}^{N}\,\exp\Bigl\{ \alpha\,{\cal R}^{\rm c}_l + \alpha\,
{\cal R}^{\rm rest}_l\Bigr\},
\eq
where $N = n(n-1)/2$ and $l$ runs over all pairings of external, charged,
fermion lines. It follows
\bqa
E(K) &=& \int\,d^4K'\,\Phi(K')
\Bigl\{ \frac{1}{(2\,\pi)^4}\,\int\,d^4x\,\exp\Bigl[ i\,\spro{(K-K')}{x} +
\alpha\,\asums{l}\,{\cal R}^{\rm c}_l\Bigr]\Bigr\},  \nl
\Phi(K) &=& \frac{1}{(2\,\pi)^4}\,\int\,d^4y\,\exp\lpar i\,\spro{K}{y}\rpar\,
\prod_{l=1}^{N}\,\exp\Bigl\{ \alpha\,{\cal R}^{\rm rest}_l\Bigr\}.
\eqa
Consider now
\bqa
{\cal H}_N\lpar K-K'\rpar &=& \frac{1}{(2\,\pi)^4}\,\int\,d^4x\,
\exp\Bigl[ i\,\spro{(K-K')}{x} +
\alpha\,\sum_{l=1}^{N}\,{\cal R}^{\rm c}_l\Bigr]  \nl
{}&=& \itpf\,\int\,d^4K_{\ssN}\,\itpf\,\int\,d^4x\,\exp\Bigl\{
i\,\spro{K_{\ssN}}{x} + \alpha\,{\cal R}^{\rm c}_{\ssN}\Bigr\}  \nl
{}&\times& \int\,d^4y\,\exp\Bigl\{ i\,\spro{(K-K'-K_{\ssN})}{y} +
\alpha\,\sum_{l=1}^{\ssN-1}\,{\cal R}^{\rm c}_l\Bigr\}.
\eqa
This result can be transformed into
\bqa
{\cal H}_N\lpar K-K'\rpar &=& \itpf\,\int\,d^4K_{\ssN}\,
\int\,d^4y\,\exp\Bigl\{ i\,\spro{(K-K'-K_{\ssN})}{y} +
\alpha\,\sum_{l=1}^{\ssN-1}\,{\cal R}^{\rm c}_l\Bigr\}  \nl
{}&\times& C_{\ssN}\,
\int_0^{\infty}\,d\sigma_{\ssN}d\sigma'_{\ssN}\,\lpar \sigma_{\ssN}
\sigma'_{\ssN}\rpar^{\alpha A_{\ssN}-1}\,\delta^4\lpar
\sigma_{\ssN}p_{i_N} + \sigma'_{\ssN}p_{j_N} - K_{\ssN}\rpar.
\eqa
with
\bq
C_{\ssN} = \frac{1}{\egams{\alpha A_{\ssN}}}.
\eq
This equation can be written as
\bqa
{\cal H}_N\lpar K-K'\rpar &=& C_{\ssN}\,\int_0^{\infty}\,d\sigma_{\ssN}
d\sigma'_{\ssN}\,\lpar \sigma_{\ssN}\sigma'_{\ssN}\rpar^{\alpha A_{\ssN}-1}\,
{\cal H}_{N-1}\lpar K-K' - \sigma_{\ssN}p_{i_N} - 
\sigma'_{\ssN}p_{j_N}\rpar,  \nl
{\cal H}_0(K) &=& \delta^4(K).
\eqa
The process can be iterated until we reach the final result
\bq
{\cal H}_N\lpar K-K'\rpar = \prod_{l=1}^{N}\,
C_l\,\int_0^{\infty}\,d\sigma_ld\sigma'_l\,\lpar \sigma_l\sigma'_l
\rpar^{\alpha A_l - 1}\,
\delta^4\lpar K - K' - \sum_{l=1}^{N}\lpar \sigma_l p_{i_l} +
\sigma'_l p_{j_l}\rpar\rpar.
\eq
Define
\bq
x_k = \sum_l\,\lpar \sigma_l + \sigma'_l\rpar|_{i_l,j_l \in \{k\}},
\qquad 1 \le k \le n,
\eq
where $n$ is the number of emitters, $N$ the corresponding number of pairs
and $\{k\}$ is the set of pairs that have one $k$-line. Then
\bq
\sum_{l=1}^{N}\lpar \sigma_l p_{i_l} + \sigma'_l p_{j_l}\rpar =
\sum_{k=1}^{n} x_k\,p_k.
\eq
We recall that \eqn{gir} is valid only for negative exponents,
i.e.\ for positive $A_{ij}$. Since some of the $A_{ij}$ may be negative,
depending on the product $\theta_i\theta_j$, we are forced to consider
an alternative derivation. Starting from
\bqa
E(K) &=& \int\,d^4K'\,\Phi(K')\,{\cal H}(K-K'),  \nl
{\cal H}(K) &=& \itpf\,\int\,d^4x\,\exp\Bigl\{ i\,\spro{K}{x} +
\alpha\asums{l}\,{\cal R}^{\rm c}_l\Bigr\},
\eqa
we derive
\bq
{\cal H}(K) = \itpf\,\prod_{l=1}^{N}\,\Bigl[ e^{-\Delta^{\rm c}_{\rm IR}}\,
\frac{\tHss}{s_{ij}}\Bigr]^{-\alpha\,A_l}\,
\int\,d^4x\,\exp\lpar i\,\spro{K}{x}\rpar\,\prod_{i=1}^{n}\,
\lpar i\,\spro{p_i}{x}\rpar^{-\alpha\asums{l\in\{l_{i}\}}\,A_l}.
\eq
As before $l_i$ denotes the set of all pairs that include the $i$-line.
If the sum
\bq
A_i = \asums{l\in\{l_{i}\}}\,A_l
\eq
is positive the derivation follows as before. In the case of $A_i$ negative
we have to consider another integral representation,
\bq
\lpar i\,\spro{p_i}{x}\rpar^{-\alpha A_i} = \frac{1}{\egam{\alpha A_i}}\,
\int_0^{\infty}\,d\sigma\,\sigma^{\alpha A_i - 1}\,\Bigl[ \exp\lpar -\,i 
\spro{\sigma p_i}{x}\rpar - 1\Bigr],
\eq
which is valid for
\bq
- 1 < \Reb \lpar \alpha A_i \rpar < 0.
\eq
By standard arguments it follows
\bqa
{\cal H}(K) &=& \prod_{l=1}^{N}\,\Bigl[ e^{-\Delta^{\rm c}_{\rm IR}}\,
\frac{\tHss}{s_{ij}}\Bigr]^{-\alpha\,A_l}  \, e^{\alpha\delta_{ij}}\,
\prod_{i=1}^{n}\,\int_0^{\infty}\,d\sigma_i\,
\sigma_i^{\alpha\asums{l\in\{l_{i}\}}\,A_l - 1}  \nl
{}&\times& \Bigl[
1 - \theta\lpar -\asums{l\in\{l_{i}\}}\,A_l\rpar\,{\cal P}(\sigma_i)\Bigr]
\,\delta^4\lpar \sum_{i=1}^{n}\,\sigma_i p_i - K\rpar,
\eqa
where we have introduced a projector ${\cal P}$
\bq
{\cal P}(\sigma_i)\,\delta^4\lpar \sum_{j=1}^{n}\,\sigma_j p_j - K\rpar =
\delta^4\lpar \sum_{j\ne i}\,\sigma_j p_j - K\rpar.
\eq
\subsection{The case $2 \to 2$}
For a $2 \to 2$ process we have $n = 4$ external particles and $N = 6$
pairs of emitters. Let us assume that all exponent are positive and
consider
\bq
{\cal H}_6 = \prod_{l=1}^{6}\,C_l\,
\int_0^{\infty}\,d\sigma_ld\sigma'_l\,\lpar \sigma_l\sigma'_l
\rpar^{\alpha A_l - 1}\,  
\delta^4\lpar K - K' - \sum_{l=1}^{N}\lpar \sigma_l p_{i_l} +
\sigma'_l p_{j_l}\rpar\rpar.
\eq
With the following identification
\bqa
\ba{ccc}
i - \mbox{line} & j - \mbox{line} & l - \mbox{pair} \\
1 & 2 & 1 \\
1 & 3 & 2 \\
1 & 4 & 3 \\
2 & 3 & 4 \\
2 & 4 & 5 \\
3 & 4 & 6 \\
\ea
\eqa
we obtain
\bqa
\sum_{l=1}^{6} \lpar \sigma_l p_{i_l} + \sigma'_l p_{j_l}\rpar &=&
\lpar \sigma_1 + \sigma_2 + \sigma_3\rpar\,p_1 + \lpar \sigma'_1 +
\sigma_4 + \sigma_5\rpar\,p_2  \nl
{}&+& \lpar \sigma'_2 + \sigma'_4 + \sigma_6\rpar\,p_3 +
\lpar \sigma'_3 + \sigma'_5 + \sigma'_6\rpar\,p_4.
\eqa
Therefore the object to compute is
\bqa
{\cal R}_6 &=& \int_{-\infty}^{\infty}\,\prod_{i=1}^{4}\,dx_i\,
\int_0^{+\infty}\,\prod_{l=1}^{6}\,d\sigma_ld\sigma'_l\,
C_l\,\lpar \sigma_l\sigma'_l\rpar^{\alpha A_l - 1}\,
\delta^4\lpar K - K' - \sum_{i=1}^{4}\,x_ip_i\rpar  \nl
{}&\times& \delta\lpar x_1 - \sigma_1-\sigma_2-\sigma_3\rpar\,
\delta\lpar x_2 - \sigma'_1-\sigma_4-\sigma_5\rpar \nl
{}&\times& \delta\lpar x_3 - \sigma'_2-\sigma'_4-\sigma_6\rpar\,
\delta\lpar x_4 - \sigma'_3-\sigma'_5-\sigma'_6\rpar.
\eqa
With ${\cal R}_6$ expressed as the integral of some $\Sigma_6$,
\bq
{\cal R}_6 = \int_{-\infty}^{\infty}\,\prod_{i=1}^{4}\,dx_i\,
\delta^4\lpar K - K' - \sum_{i=1}^{4}\,x_ip_i\rpar\,\Sigma_6,
\eq
we perform the various $\sigma$ integrations, starting with the trivial
ones,
\bqa
\Sigma_6 &=& \int_0^{+\infty}\,\prod_{l=2}^{6}\,d\sigma_l\,\prod_{l'=4}^{6}\,
\lpar x_1-\sigma_2-\sigma_3\rpar^{\alpha A_1 - 1}\,
\lpar x_2-\sigma_4-\sigma_5\rpar^{\alpha A_1 - 1}  \nl
{}&\times& \sigma_2^{\alpha A_2 - 1}\,\lpar x_3-\sigma'_4-\sigma_6
\rpar^{\alpha A_2 - 1}\,
\sigma_3^{\alpha A_3 - 1}\,\lpar x_4-\sigma'_5-\sigma'_6
\rpar^{\alpha A_3 - 1} \nl
{}&\times& \sigma_4^{\alpha A_4 - 1}\,\lpar\sigma'_4\rpar^{\alpha A_4 - 1}\,
\sigma_5^{\alpha A_5 - 1}\,\lpar\sigma'_5\rpar^{\alpha A_5 - 1}\,
\sigma_6^{\alpha A_6 - 1}\,\lpar\sigma'_6\rpar^{\alpha A_6 - 1}.
\eqa
After integration a set of consistency conditions will emerge,
\bqa
x_1 &\ge& \sigma_2+\sigma_3, \qquad x_2 \ge \sigma_4+\sigma_5,  \nl
x_3 &\ge& \sigma'_4+\sigma_6, \qquad x_4 \ge \sigma'_5+\sigma'_6.
\eqa
The next step is to perform the $\sigma_2$-integration,
\bq
\int_0^{\infty}\,d\sigma_2\,\lpar\sigma_2\rpar^{\alpha A_2 - 1}\,
\lpar x1-\sigma_3-\sigma_2\rpar^{\alpha A_1 - 1} =
\ebe{\alpha A_1}{\alpha A_2}\, \lpar x_1-\sigma^3\rpar^{\alpha
(A_1+A_2) - 1},
\eq
which requires $x_1 \ge \sigma_3$ and where $B$ is the Euler's beta-function.
Next we integrate over $\sigma_3$,
\bq
\ebe{\alpha A_1}{\alpha A_2}\,\int_0^{\infty}\,d\sigma_3\,
\lpar\sigma_3\rpar^{\alpha A_3 - 1}\,\lpar x_1 - \sigma_3
\rpar^{\alpha(A_1+A_2) - 1} 
= {{\prod_{i=1,3}\,\egam{\alpha A_i}}\over 
{\egam{\alpha\sum_{i=1,3} A_i}}}\, \lpar x_1\rpar^{\alpha\sum_{i=1,3} A_i -
1}.
\eq
Also the $\sigma_4, \sigma_5$ integrations give
\bq
\int_0^{\infty}\,d\sigma_4\,\lpar\sigma_4\rpar^{\alpha A_4 - 1}\,
\lpar x_2-\sigma_5-\sigma_4\rpar^{\alpha A_1 - 1} =
\ebe{\alpha A1}{\alpha A_4}\,\lpar x_2-\sigma_5\rpar^{\alpha(A_1+A_4) - 1},
\eq
\bqa
{}&{}&\ebe{\alpha A_1}{\alpha A_4}\,\int_0^{\infty}\,d\sigma_5\,
\lpar\sigma_5\rpar^{\alpha A_5 - 1}\,\lpar x_2-\sigma_5\rpar^{\alpha(A_1+A_4) -
1} \nl
{}&=& {{\egam{\alpha A_1}\egam{\alpha A_4}\egam{\alpha A_5}}\over
{\egam{\alpha\lpar A_1+A_4+A_5\rpar}}}\,\lpar x_2\rpar^{\alpha(A_1+A_4+A_5) -
1},
\eqa
Similar results hold for the $\sigma_6, \sigma'_4$ and $\sigma'_5, \sigma'_6$
integration, with a total result
\bqa
\Sigma_6 &=& {{\prod_{l=1}^{6}\,\egams{\alpha A_l}}\over
{\prod_{i=1}^{4}\,\egam{\alpha A_{\{i\}}}}}\,\prod_{i=1}^{4}\,
\lpar x_i\rpar^{\alpha A_{\{i\}} -  1},  \nl
A_{\{1\}} &=& A_1+A_2+A_3, \qquad A_{\{2\}} = A_1+A_4+A_5,  \nl
A_{\{3\}} &=& A_2+A_4+A_6, \qquad A_{\{4\}} = A_3+A_5+A_6,
\eqa
which can be easily generalized to
\bq
\Sigma_{\ssN} = \prod_{l=1}^{N}\,\egams{\alpha A_l}\,\Bigl[
\prod_{i=1}^{n}\,\egam{\alpha\,\sum_{k\in \{k_i\}} A_k} \Bigr]^{-1}\,
\prod_{i=1}^{n}\,\lpar x_i\rpar^{\alpha\,\sum_{k\in \{k_i\}} A_k - 1},
\eq
where, for $i$ fixed, the index $k \in \{k_i\}$ runs over all pairs that 
contain the line $i$ and $\forall i, x_i \ge 0$. 
\section{Combining real and virtual corrections\label{sectIR}}
The emission of a real photon from the $ij$ pair is described by a
structure function language with an exponent $\alpha A_{ij}-1$, with
\bq
A_{ij} = \frac{\theta_i\theta_j}{\pi}\,\Bigl[ 1 - \rho_{ij}\,
\frac{\spro{p_i}{p_j}}{m^2_j-\rho^2_{ij}\,m^2_i}\,\ln\frac{\rho^2_{ij}\,m^2_i}
{m^2_j}\Bigr].
\eq
For virtual photons the overall, universal, factor 
\bq
B = -\,\frac{i}{8\,\pi^2}\,\asums{i<j}\,\theta_i\theta_j\,B_{ij},
\eq
is exponentiated and the relevant quantity is $\exp\lpar 2\,\alpha\,
\Reb B\rpar$. For real radiation we have another overall exponentiation where 
the infrared-divergent object, for reach pair, is given in \eqn{strf},
\bq
\Bigl[ e^{-\Delta^{\rm c}_{\rm IR}}\,\frac{\tHss}
{s_{ij}}\Bigr]^{-\alpha A_l}\,e^{\alpha\delta_{ij}} = 
\exp\Bigl\{ \alpha \,A_{ij}\,\Bigl[
\Delta^{\rm c}_{\rm IR} - \ln\frac{\tHss}{s_{ij}}\Bigr] + \alpha\,\delta_{ij}
\Bigr\},
\eq
where $l = \{ij\}$ is the emitting pair that we are considering. Here
\bq
\Delta^{\rm c}_{\rm IR} = \Ddrh - 2\,\gamma + \frac{3}{2}, \qquad
\delta_{ij}= -\,\frac{\theta_i\theta_j}{\pi}\,\lpar \frac{1}{2} + 
\frac{\pi^2}{6} \rpar
\eq
Structure function language means that the overall, real + virtual, exponent 
is multiplied by
\bq
\prod_i\,\frac{\beta_i\,\lpar x_i\rpar^{\beta_i-1}}{\egam{\beta_i+1}},
\eq
where $i$ runs over external charged lines. Moreover
\bq
\beta_i = \alpha\,\sum_{k \in k_i}\,A_k,
\eq
where the sum is limited to those pairs containing $i$. 
\subsection{IR finite exponent}
Once virtual and real exponentiation are combined we have cancellation of the 
infrared pole and some global remainder that reads as follows:
\bqa
\frac{\alpha}{\pi}\,\theta_i\theta_j\,{\cal F}_{ij}  &=&
\alpha\,\Bigl\{A_{ij}\,\Bigl[ \Delta^{\rm c}_{\rm IR} + \ln\frac{s_{ij}}{\tHss}
\Bigr] - \frac{\theta_i\theta_j}{\pi}\,\lpar\frac{1}{2} +
\frac{\pi^2}{6}\rpar\Bigr\}
- \frac{i\,\alpha}{4\,\pi^2}\,\theta_i\theta_j\,\Reb B_{ij}  \nl
{}&=& \frac{\alpha\,\theta_i\theta_j}{\pi}\,\Bigl\{\Bigl[
1 - \rho_{ij}\,\frac{\spro{p_i}{p_j}}{m^2_j-\rho^2_{ij}\,m^2_i}\,
\ln\frac{\rho^2_{ij}\,m^2_i}{m^2_j}\Bigr]\,\lpar 
\Delta^{\rm c}_{\rm IR} + \ln\frac{s_{ij}}{\tHss}\rpar  \nl
{}&-& \frac{1}{2} - \frac{\pi^2}{6} + \Reb B^{\rm IR}_{ij}\,\Ddrh + 
\Reb B^{\rm fin}_{ij}\Bigr\},
\eqa
where $1/\ept$ shows the virtual infrared pole. The residue and the finite part
of the virtual corrections are
\bqa
B^{\rm IR}_{ij} &=& - 1 + \epsilon_i\epsilon_j\,\spro{p_i}{p_j}\,F^{ij}_1,  \nl
B^{\rm fin}_{ij} &=& -\,\ln\frac{m_im_j}{\tHss} + \epsilon_i\epsilon_j\,
\spro{p_i}{p_j}\,\Bigl[ F_1\,\ln\frac{Q^2-i\epsilon}{\tHss} + F^{\rm rest}_2
\Bigr] + \frac{1}{2}\,F_3.
\eqa
Here $F_1$ is expressed as
\bqa
Q &=& \epsilon_i\,p_i + \epsilon_j\,p_j,  \nl
y_{1,2} &=& \frac{1}{2\,Q^2}\,\Bigl[ Q^2 + m^2_j - m^2_i \pm
\lambda^{1/2}\lpar -Q^2,m^2_i,m^2_j\rpar\Bigr],  \nl
F^{ij}_1 &=& \frac{1}{Q^2\lpar y_1-y_2\rpar}\,\Bigl[
\ln\lpar 1 - \frac{1}{y_2}\rpar - \ln\lpar 1 - \frac{1}{y_1}\rpar\Bigr].
\eqa
Furthermore introduce a shorthand notation for the K\"allen's function,
\bq
\lambda\lpar - Q^2,m^2_i,m^2_j\rpar = \Lambda^2.
\eq
We see from \eqn{defrho} that $\rho_{ij}$ is also a solution of the equation
\bq
\lpar \rho_{ij}\,p_i - p_j\rpar^2 = 0. 
\eq
If $\epsilon_i\epsilon_j = +1$ we find
\bq
\rho_{ij} = \frac{1}{2\,m^2_i}\,\Bigl[ - Q^2 - m^2_i - m^2_j + \Lambda\Bigr].
\eq
If instead $\epsilon_i\epsilon_j = -1$ we have
\bq
\rho_{ij} = \frac{1}{2\,m^2_i}\,\Bigl[ Q^2 + m^2_i + m^2_j + \Lambda\Bigr].
\eq
In both cases we derive a noticeable relation,
\bq
\frac{\rho}{m^2_j-\rho^2m^2_i} = - \frac{1}{\Lambda}.
\eq
Moreover, it is straightforward to show that
\bq
\frac{1}{Q^2\lpar y_1-y_2\rpar} = \frac{1}{\Lambda}.
\eq
Consider now the quantity $Y$ defined as
\bq
Y = \frac{y_1\lpar y_2-1\rpar}{y_2\lpar y_1-1\rpar}.
\eq
It follows that $Y$ can be expressed as
\bq
Y = \frac{\lpar Q^2+m^2_i+m^2_j+\Lambda\rpar^2}{4\,m^2_im^2_j} =
\frac{4\,m^2_im^2_j}{\lpar Q^2+m^2_i+m^2_i-\Lambda\rpar^2}.
\eq
Similarly we obtain
\[
\frac{\rho^2_{ij}m^2_i}{m^2_j} =
\left\{
\begin{array}{ll}
\frac{\lpar Q^2+m^2_i+m^2_j-\Lambda\rpar^2}{4\,m^2_im^2_j} & \mbox{if
$\epsilon_i\epsilon_j = +1$} \\
\frac{\lpar Q^2+m^2_i+m^2_j+\Lambda\rpar^2}{4\,m^2_im^2_j} & \mbox{if
$\epsilon_i\epsilon_j = -1$} \\
\end{array}
\right.
\]
In other words, an important result can be derived, namely
\bq
\ln\frac{\rho^2_{ij}\,m^2_i}{m^2_j} = - \epsilon_i\epsilon_j\,\ln Y, \qquad
Y = \frac{y_1\lpar y_2-1\rpar}{y_2\lpar y_1-1\rpar}.
\eq
The IR-finite exponent is therefore $\alpha/\pi\,\theta_i\theta_j\,
{\cal F}_{ij}$, with
\bqa
{\cal F}_{ij} &=& 1 - 2\,\gamma - \frac{\pi^2}{6} + 
\ln\frac{s_{ij}}{m_im_j} +
\frac{1}{2}\,\Reb F_3 + \epsilon_i\epsilon_j\,\frac{\spro{p_i}{p_j}}{\Lambda}\,
\Bigl\{\Bigl[ \ln\frac{\mid Q^2\mid}{s_{ij}} \nl
{}&-& \frac{3}{2} + 2\,\gamma\Bigr]\,\ln Y + \Reb f^{\rm rest}_2\Bigr\},
\label{naexp}
\nl
\eqa
and with
\bq
L_{ij} = \ln\frac{\mid Q^2\mid}{m_im_j}, \qquad
F^{\rm rest}_2 = \frac{1}{\Lambda}\,f^{\rm rest}_2.
\eq
\subsection{Asymptotic limits and general considerations}
\begin{itemize}
\item[a)] {\em the case} $Q^2 \gg m^2$
\end{itemize}
It is important to show the asymptotic behavior of this exponent in the region
$\mid Q^2\mid \gg m^2_i,m^2_j$.
We easily obtain that
\bq
\Lambda \sim -\,Q^2, \qquad y_1 \sim -\frac{m^2_i}{Q^2}, \quad
y_2 \sim 1 - \frac{m^2_j}{Q^2},
\eq
giving the asymptotic limit of $Y$ as 
\bq
Y \sim \frac{m^2_im^2_j}{(Q^2)^2}.
\eq
Using the asymptotic behavior of \eqn{asympt} and also \eqn{sasym} we derive
\bq
{\cal F}_{ij} \sim 2\,\gamma\,\lpar L_{ij} - 1 \rpar - \frac{3}{2}\,L_{ij} +
2 - \frac{\pi^2}{3},
\label{norma}
\eq
which shows, among other things, a cancellation of the $\ln^2$ terms.

We can easily check that the exponent $\beta_{ij} = \alpha\,A_{ij} - 1$
has the usual asymptotic behavior. With
\bq
s = - \lpar p_i + p_j\rpar^2, \qquad m_i = m_j = m, \qquad 
\theta_i\theta_j = -1,
\eq
in the limit $s \gg m^2$ we get
\bq
\alpha\,A_{ij} \sim \frac{\alpha}{\pi}\,\lpar \ln\frac{s}{m^2} - 1\rpar.
\eq
The result of \eqn{norma} explains the normalization in the definition of the 
coplanar factor. The infrared finite overall exponent has, in the asymptotic
region $\mid Q^2\mid \gg m^2$, the correct behavior to reproduce the
exponentiation commonly employed to describe initial state radiation
in $e^+e^-$-annihilation, at least up to terms $\ord{\alpha^2}$ and
without hard photons,
\bqa
G(x) &=& \frac{\beta}{\egam{\beta+1}}\,x^{\beta-1}\,\exp\Bigl\{
-\beta\gamma + \delta^{\rm{V+S}}\Bigr\},  \nl
\delta^{\rm{V+S}} &=& \frac{\alpha}{\pi}\,\lpar \frac{3}{2}\,
\ln\frac{s}{\mes} - 2 + \frac{\pi^2}{3}\rpar, \qquad \beta = 
\frac{2\,\alpha}{\pi}\,\lpar \ln\frac{s}{\mes} - 1\rpar.
\label{srad}
\eqa
In the above result $G$ is the so-called radiator function which is connected
to structure functions $D$ by the relation
\bq
G(x) = \int_x^1\,dz\,D(z)\,D(\frac{x}{x}).
\eq
Note that \eqn{srad} is sometimes written as
\bqa
G(x) &=& \frac{\beta}{\egam{\beta+1}}\,x^{\beta-1}\,\exp\Bigl\{
-\beta\gamma + \delta^{\rm YFS}\Bigr\}\,\lpar 1 + \delta_{\rm S}\rpar,  \nl
\delta^{\rm YFS} &=& \frac{\alpha}{\pi}\,\lpar \frac{1}{2}\,\ln\frac{s}{\mes} -
1 + \frac{\pi^2}{3}\rpar,  \nl
\delta_{\rm S} &=& \frac{1}{2}\,\beta + \frac{1}{2}\,
\lpar \frac{\alpha}{\pi}\rpar^2\,\ln^2\frac{s}{\mes},
\eqa
a form which follows from the evaluation of the YFS soft form-factor
\bq
\beta\,\ln\varepsilon + 2\,\delta_{\rm YFS},
\eq
$\varepsilon$ being the parameter introduced to limit the multiplicity of very
soft photons.

Moreover, if we neglect constant terms, our exponent of \eqn{naexp} reproduces 
the leading behavior of the Gribov-Lipatov solution~\cite{gls} of the 
evolution equation for the electron structure function,
\bq
D(x) = \frac{x^{\eta/2-1}}{\egam{\eta/2}}\,\exp\Bigl\{\frac{\eta}{4}\,\lpar 
\frac{3}{2} - 2\,\gamma\rpar\Bigr\}, \qquad
\eta = -6\,\ln\lpar 1 - \frac{\alpha}{3\,\pi}\,\ln\frac{s}{\mes}\rpar,
\eq
which is valid in the soft limit. The factor $\delta_{ij}$ inserted in 
\eqn{defcop} has the purpose of reproducing the constant terms of
\eqn{srad}, therefore increasing the accuracy of the approximation.

In conclusion the coplanar approximation, in the limit $x_i \ll 1, \forall i$
coincides precisely with the exact expression resulting from the soft-photon
re-summation, as given for $e^+e^- \to \ph^*$ in the classic YFS treatment.
We want to stress that, for a general process, there are ambiguities in the
choice of overall exponent in \eqn{naexp}. Indeed the asymptotic factor
\bq
- \frac{3}{2}\,L_{ij} + 2 - \frac{\pi^2}{3},
\eq
which is sub-leading, is tailored to reproduce the asymptotic form of the
soft + virtual one loop corrections to the vertex $e^+e^-\gamma$, i.e.\
$2\,\Reb F_{\rm dirac} + \delta_{\rm soft}$. Single logarithms and
constant terms in the overall exponent can never be exact, unless the
full virtual + hard part of the spectrum is included. Therefore, the
accuracy of the result is always limited to
\bq
\beta\,x^{\beta-1}\,\Bigl[ 1 + \ord{\alpha\,L_{\rm coll}}\Bigr].
\eq
As mentioned in \sect{bckg} the missing parts of the hard photon spectrum
and of virtual corrections violate the well-known KLN result that the 
inclusive corrections are always small and free of large logarithms for a 
pair of final state emitters.
Therefore, the accuracy of our result is, in this case, controlled only if 
tight cuts are imposed on the invariant masses of the final state pairs.
To give a concrete example the cross-section for $e^+e^- \to \barf f$
that includes exact $\ord{\alpha}$ final state radiation 
is~\cite{Montagna:1993mf}
\bqa 
\sigma_c\lpar\sman\rpar &=&  
\frac{\alpha}{4\pi}\qfs\sigma^0\lpar{\sman}\rpar
\biggl\{
-2\lpar1-\zvar\rpar^2 
+4\biggl[
\lpar\zvar+\frac{\zvars}{2}+2\ln\lpar 1-\zvar\rpar\rpar\ln\frac{\sman}{\mfs}
\\{}&&
+\zvar\lpar 1 +\frac{\zvar}{2}\rpar\ln\zvar 
+2\ztwo-2\li{2}{1-\zvar}
-2\ln\lpar 1-\zvar\rpar + \frac{5}{4}- 3\zvar - \frac{\zvars}{4}   
\biggr]
\biggr\},
\label{imc1}
\nn
\eqa
where $\zvar=\Mlones\lpar\ff\barf\rpar/\sman$. The difference between the factor
\bq
\ln\lpar 1 - z\rpar\,\Bigl[ \ln\frac{\sman}{\mfs} - 1\Bigr],
\eq
which is then exponentiated and the full result of \eqn{imc1} is due to
hard, virtual and real, photons and is responsible for the correct limit
$\zvar \to 0$,
\bq
\sigma_c = \sigma^{0}\lpar{\sman}\rpar
\lpar 1 + \frac{3}{4}\frac{\alpha}{\pi}\qfs\rpar,
\label{FSR_nocut}
\eq
\begin{itemize}
\item[b)] {\em the case} $Q^2 \ll m^2$
\end{itemize}
Finally we consider the case of one incoming/ one outgoing electron. For
\bq
t = - \lpar p_i - p_i\rpar^2, \qquad - t \ll \mes,
\eq
we obtain
\bq
\rho_{ij} \sim 1 + \lpar -\frac{t}{\mes}\rpar^{1/2}, \qquad
\alpha\,A_{ij} \sim \frac{\alpha}{\pi}\,\lpar -\frac{t}{\mes}\rpar^{1/2},
\eq
showing a power law behavior in the exponent, to be compared with the
logarithmic one for $-t \gg \mes$.

When $Q^2 \ll m^2_i,m^2_j$, we derive
\bq
\Lambda \sim \lpar m^2_i - m^2_j\rpar^2, \qquad \rho_{ij} \sim  
\frac{m^2_i+m^2_j}{2\,m^2_i}.
\eq
Therefore the real emission is controlled by a coefficient
\bq
{\cal A}_{ij}  = 
1 - \rho_{ij}\,\frac{\spro{p_i}{p_j}}{m^2_j-\rho^2_{ij}\,m^2_i}\,
\ln\frac{\rho^2_{ij}\,m^2_i}{m^2_j} \sim
1 + 2\,\frac{(1+r)^2}{4-r^2(1+r)^2}\,\ln\frac{r(1+r)}{2},
\eq
with $r = m^2_i/m^2_j$, which, for equal masses, reproduces the correct limit
\bq
{\cal A}_{ij}\lpar Q^2,m_i=m_j=m\rpar \to 0, \qquad \mbox{for}\quad 
Q^2 \to 0,
\eq
where a non-accelerated charge does not radiate.
From these result we see that the normalization of the coplanar factor in 
\eqn{defcop} has been chosen to avoid appearance of spurious logarithms,
$\ln(Q^2/m^2)$, for $Q^2 \ll m^2$.
One may wonder whether the limit of very small momentum-transfer is
relevant for any physical situation. Consider the process
$e^+(p_+) + e^-(p_-) \to e^-(q_-) + X(q_{\ssX})$ and define
\bq
Q = p_- - q_-, \qquad y = \frac{\spro{p_+}{Q}}{\spro{p_+}{p_-}}.
\eq
In the region of forward $e^-$ scattering we have
\bq
Q^2 \ge \mes\,\frac{y^2}{1-y},
\eq
where $y$ is bounded by
\bq
\frac{M^2_0}{s} \le y \le 1 - \frac{\me}{\sqrt{s}}, \qquad
s = - \lpar p_+ + p_-\rpar^2,
\eq
and $M_0$ is the minimum invariant mass of the $X$-system. Whenever this mass 
is very low with respect to $s$ we may reach the regime $Q^2 \ll \mes$.

As a final consideration, note that we are treating all charged fermionic
lines on the same footing since only this combination of radiation is a
meaningful gauge-invariant concept. For a general process, therefore, the
interference between different legs is a fundamental part of the QED
corrections and not an additional minor effect. Interference must be 
meaningfully definable, in particular when one exponentiates, see a 
discussion in ref.~\cite{Kleiss:1992ch}. In certain
situations interference is responsible for changing the scale in the 
leading logarithms but it should be present, as a matter of principle, also
for those situations, as in forward scattering, where the typical scale
is not large with respect to fermion masses. Strictly speaking the
exponentiation of `soft' initial-final interference is not accurate enough and,
therefore, insufficient to describe annihilation processes around resonances. 
We will not dwell upon this subject any longer and refer to~\cite{cssf}.
\section{Further refinements\label{refin}}
There are several reasons to increase the value of our approximation and
,for the sake of simplicity, we start our considerations by examining the case 
of a $2 \to 2$ process. The result may be cast into the following form,
\bqa
\sigma &\propto& \int\,dPS_2\,\int\,d^4K\,\Phi(K')\,\int\,\prod_{i=1}^{4}\,dx_i
\nl
{}&\times&
\delta^4\lpar p_+ + p_- - q_+ - q_- - K' - \asums{i}\,x_ip_i\rpar\,
\prod_{i}\,\frac{\beta_i}{\egam{1+\beta_i}}\,
\lpar x_i\rpar^{\beta_i - 1}\,\beta_0,
\label{exa}
\eqa
where $\beta_0$ is the Born matrix element, $\beta_i = \alpha A_i$ and
\bq
p_1 = p_+, \quad p_2 = p_-, \quad p_3 = q_-, \quad p_4 = q_+.
\eq
Furthermore $dPS_2$ is the two-body phase-space.
If we neglect terms of $\ord{\alpha}$ in the flux-function, i.e.\
$\Phi(K') = \delta^4(K')$, the argument of the delta-function in \eqn{exa}
becomes
\bq
\lpar 1-x_1\rpar\,p_+ + \lpar 1-x_2\rpar\,p_- - \lpar 1+x_3\rpar\,q_- -
\lpar 1+x_4\rpar\,q_-.
\eq
Next we introduce scaled momenta,
\bqa
{\hat p}_+ &=& \lpar 1-x_1\rpar\,p_+, \qquad
{\hat p}_- = \lpar 1-x_2\rpar\,p_-, \nl
{\hat q}_+ &=& \lpar 1+x_4\rpar\,q_+, \qquad
{\hat q}_- = \lpar 1+x_3\rpar\,q_-,
\eqa
and derive four-momentum conservation in terms of the radiative process
which incorporates the photons emitted along the directions of the charged 
fermions. There seems to be a clash between kinematics and matrix element;
the original procedure requires a reorganization of the perturbative
expansion which starts with the matrix element in soft approximation while
the delta-function expressing conservation is kept exact, transformed with
other ingredients into the photon spectral function which is again approximated
to introduce conservation at the level of scaled momenta, as it would appear
in the structure function language. In any intermediate step we are not
authorized to use energy-momentum conservation since there is no
delta-function to use.

The lowest order factor in the perturbative expansion of the squared matrix 
element, $\beta_0$, is however constructed with non-scaled momenta. Typically, 
we will have
\bq
\beta_0 \propto \spro{q_-}{p_-}\,\spro{q_+}{p_+} + \spro{q_-}{p_+}\,
\spro{q_+}{p_-}.
\eq
Let us change variables according to
\bqa
x_+ &=& 1 - x_1, \qquad x_- = 1 - x_2,  \nl
\frac{1}{y_+} &=& 1 + x_4, \qquad \frac{1}{y_-} = 1 + x_3,  
\eqa
so that the set of momenta satisfying conservation is
\bqa
{\hat p}_+ &=& x_+\,p_+, \qquad {\hat p}_- = x_-\,p_-,  \nl
{\hat q}_+ &=& \frac{q_+}{y_+}, \qquad {\hat q}_- = \frac{q_-}{y_-},  \nl
\prod_{i=1}^{4}\,dx_i &=& \frac{1}{y^2_+y^2_-}\,dx_+dx_-dy_+dy_-.
\eqa
As a consequence the Born matrix element can be cast into the following form:
\bq
\beta_0 \propto \frac{y_+y_-}{x_+x_-}\,\Bigl[
\spro{{\hat p}_-}{{\hat q}_-}\,\spro{{\hat p}_+}{{\hat q}_+} +
\spro{{\hat p}_+}{{\hat q}_-}\,\spro{{\hat p}_-}{{\hat q}_+}\Bigl].
\eq
Futhermore,$\beta_0$ will contain an overall factor $s^{_3}$ from the 
$s$-channel propagator and from the root of the K\"allen function.
Given the relation between $s$ and $\shat$,
\bq
s = - \lpar p_+ + p_-\rpar^2 = \frac{\shat}{x_+x_-},
\eq
we obtain
\bq
\beta_0 = \frac{\pi}{2\,\shat}\int\,d\that\,\beta^{\rm inv}_0, 
\eq
and, after transforming variables,
\bq
\prod_{i=1}^{4}\,\int\,dx_i\,\beta^{\rm inv}_0 = 
\int\, dx_+dx_-dy_+dy_-\,\frac{x^2_+x^2_-}{y_+y_-}\,\frac{\that^2+\uhat^2}
{\shat^3}.
\eq
In deriving this result we have used scaled invariants, defined by
\bqa
\shat &=& - \lpar {\hat p}_+ + {\hat p}_- \rpar^2 =
- \lpar {\hat q}_+ + {\hat q}_- \rpar^2,  \nl
\that &=& - \lpar {\hat p}_+ - {\hat q}_+ \rpar^2 =
- \lpar {\hat p}_- - {\hat q}_- \rpar^2,  \nl
\uhat &=& - \lpar {\hat p}_+ - {\hat q}_- \rpar^2 =
- \lpar {\hat q}_+ - {\hat p}_- \rpar^2. 
\eqa
From phase space considerations and from positivity for all $x_i$ it follows
that
\bq
0 \le x_{\pm},\,y_{\pm} \le 1.
\eq
As a next step, we will show that it is possible to introduce, in a way that 
is consistent with the perturbative approach, a new lowest order result: 
it contains ${\hat\beta}^{\rm inv}_0$ instead of $\beta^{\rm inv}_0$,
\bq
{\hat\beta}^{\rm inv}_0 = \frac{\that^2+\uhat^2}{\shat^3},
\eq
the difference between the two formulations being of order $\alpha$. This 
difference is known and computable so that perturbation theory indeed starts 
with a radiative kernel and the re-summation of all photons emitted along the 
directions of the external, charged, fermions.
Let us consider this difference in more detail. First of all it is zero for 
$x_+ = \dots y_- = 1$. Next consider the following function:
\bq
F\lpar z,\beta\rpar = \int_z^1\,dx\,{\cal D}_{\beta}(x)\,f(x),
\eq
where $\beta = \alpha A - 1$ and where we have also introduced the distribution
\bq
{\cal D}_{\beta} = \beta\,\lpar 1 - x\rpar^{\beta-1}.
\eq
Adding and subtracting a term we derive
\bq
F\lpar z,\beta\rpar = f(1)\,\lpar 1 - z\rpar^{\beta} + \beta\,\int_z^1\,
dx\,\frac{f(x) - f(1)}{1-x}\,\lpar 1 - x\rpar^{\beta}.
\eq
From this result the distribution can be computed. For instance, to second 
order in $\beta$, we have
\bqa
{\cal D}_{\beta} &=& \delta(x-1) + \beta\,\Bigl\{ \ln\lpar 1-z\rpar + \Bigl[
\frac{1}{1-x}\Bigr]_+\Bigr\}  \nl
{}&+& \beta^2\,\Bigl\{ \frac{1}{2}\,\ln^2\lpar 1-z\rpar + \Bigl[
\frac{\ln(1-x)}{1-x}\Bigr]_+\Bigr\} + \ord{\beta^3}.
\eqa
The `+'-distribution is defined, as usual, by its action on a generic test
function $g(x)$:
\bq
\int_z^1\,dx\,g(x)f_+(x) = \int_z^1\,dx\,\Bigl[ g(x) - g(1)\Bigr]\,f(x).
\eq
Consider now a simple example, where $f(x) = x^2\,{\hat f}(x)$; it follows
that
\bqa
F\lpar z,\beta\rpar &=& \int_z^1\,\beta\lpar 1 - x\rpar^{\beta-1}\,x^2
{\hat f}(x) = \int_z^1\,dx \beta\,\lpar 1 - x\rpar^{\beta-1}\,{\hat f}(x)  \nl
{}&\times& \Bigl\{ 1 + \beta\,\Bigl[ \frac{1}{{\hat f}(x)}\,\int_z^1\,dy\,
\lpar 1 + y\rpar\,{\hat f}(y) 
+ \lpar 1 - x^2\rpar\,\ln\lpar 1 - x\rpar\Bigr] +
\ord{\beta^2}\,\Bigr\}.
\nl
\eqa
Therefore, the perturbative expansion is controlled by the parameter 
$\beta = \alpha\,A - 1$ and we may compute the kernel cross-section with 
scaled momenta and fold it with the appropriate factors of 
$\beta(1-x)^{\beta-1}$. The difference
with a kernel cross-section computed with non-scaled momenta and the successive
application of the correct four-momentum conservation appears only at the next
order in $\beta$ and can be re-adjusted order-by-order in perturbation theory.

There are two reasons why one should rescale momenta in the kernel 
cross-section. In any process with a resonance in the annihilation channel
this procedure includes the possibility of a correct description of the 
radiative return directly in lowest order. There is more, this procedure
is sometimes requested by a correct treatment of gauge invariance. Consider, 
for instance, the process $e(p) + P(P) \to e(p') + X$. We write
\bq
d\sigma = \frac{1}{2\,\spro{P}{p}}\,\frac{e^2\,W_{\mu\nu}T^{\mu\nu}}
{(q^2)^2}\,\frac{d^4p'}{(2\,\pi)^3}\,\delta^+(p'^2+\mes)\,
\delta^4\lpar p - q - p'\rpar.
\eq
The factor $\delta^4(p-q-p')$ is successively promoted to become the spectral 
function $E(p-q-p')$ which, in coplanar approximation, generates 
\bq
E^{\rm c}\lpar p-q-p'\rpar \to \delta^4\lpar (1-\sigma)p - 
(1+\sigma')p' - q\rpar,
\eq
so that the kinematics is specified by 
\bq
q = {\hat p} - {\hat p}',  \qquad
{\hat p} = x p = \lpar 1 - \sigma\rpar p, \qquad 
{\hat p}' = \frac{p'}{y} = \lpar 1 + \sigma'\rpar p', 
\eq
while the leptonic tensor in soft approximation is extracted as
\bqa
T_{\mu\nu} &=& \frac{1}{2}\,q^2\,\drii{\mu}{\nu} + p_{\mu}p'_{nu} +
p_{\nu}p'_{\mu} = \frac{1}{2}\,q^2\,\drii{\mu}{\nu} + \frac{y}{x}\,\Bigl[
{\hat p}_{\mu}{\hat p'}_{nu} + {\hat p}_{\nu}{\hat p}'_{\mu}\Bigr]  \nl
{}&=& {\hat T}_{\mu\nu} + \lpar \frac{y}{x} - 1\rpar\,\Bigl[
{\hat p}_{\mu}{\hat p'}_{nu} + {\hat p}_{\nu}{\hat p}'_{\mu}\Bigr],
\eqa
and gauge invariance is respected only by ${\hat T}$, namely
$q_{\mu}{\hat T}^{\mu\nu} = q_{\nu}{\hat T}^{\mu\nu} = 0$.

One should remember that, at least in principle, some of the exponents 
$\beta_i$ inside \eqn{exa} could be negative. In this case we should write
\bq
\int\,\prod_{i}\,x_i^{\beta_i-1}\,\Bigl[ 1 - \theta(-\beta_i)\,{\cal P}(x_i)
\Bigr]\,{\hat\beta}_0,
\eq
which, by setting some of the $x_i$ to zero in the integrand, introduces the 
opportune subtraction on the original result. After changing variables, we
obtain
\bqa
{}&{}& \int_0^1\,dx_+dx_-dy_+dy_-\,\lpar 1 - x_+\rpar^{\beta_1-1}\,
\lpar 1 - x_-\rpar^{\beta_2-1}\,\lpar \frac{1}{y_-} - 1\rpar^{\beta_3-1}\,
\lpar \frac{1}{y_+} - 1\rpar^{\beta_4-1}  \nl
{}&\times& \frac{x^2_+x^2_-}{y_+y_-}\,
\Bigl[ 1 - \theta(-\beta_1)\,{\cal P}(x_+ - 1)\Bigr]\,\dots\,
\Bigl[ 1 - \theta(-\beta_4)\,{\cal P}(y_+ - 1)\Bigr]\,{\hat\beta}^{\rm inv}_0.
\nl
\eqa
Each subtraction sets one parameter to one and corresponds, figuratively
speaking, to disallow radiation from the corresponding leg.

Furthermore, one should also take into proper account the kinematical 
constraints on the process. We must require that the invariant mass of the 
outgoing fermion -- anti-fermion pair be at least $2\,m_f$ which implies
\bq
x_+x_-y_+y_- \ge 4\,\frac{m^2_f}{s}, \qquad s = - \lpar p_+ + p_- \rpar^2,
\eq
and $x_{\pm}, y_{\pm}$ cannot be zero for non-zero fermion masses.
\section{A strategy for computing $\beta_1$\label{astrat}}
The real improvement upon the present implementation of QED radiative
corrections in generic $2 \to n$ processes requires to go beyond
$\beta_0$ in \eqn{impro}. Therefore, to go beyond the present approximation 
one has to compute $\beta_1$ in \eqn{impro} or, at least to include the 
collinear singularity of the hard photon. Consider once more the process
\bq 
e^+e^- \to n\,f + \ph,
\eq
and let 
\bq
M = M_{\mu}\epsilon^{\mu}(k), \qquad M_0 \equiv M\lpar e^+e^- \to n\,f\rpar.
\eq
Let $i$ be an external fermion, for instance outgoing. Then
\bq
M_{\mu} = i\,e\,\baru(p_i)\,\lpar T^i_{\mu} + R^i_{\mu}\rpar,  \qquad
T^i_{\mu} = - i\,\frac{Q_i}{2\,\spro{p_i}{k}}\,\gadu{\mu}\,
\lpar \sla{p_i} + \sla{k}\rpar\,T^i(p_i+k),
\eq
where, for simplicity we have assumed massless fermions. $T$ represents the
contribution where the photon is emitted by the $p_i$-line with residual
amplitude $T^i(p_i+k)$ and $R$ represents the rest. Gauge invariance requires
$\spro{k}{M} = 0$ or
\bq
\baru(p_i)\,\spro{k}{R^i} = i\,Q_i\,\baru(p_i)\,T^i\lpar p_i+k\rpar.
\eq
Consider vectors $Q,n$ and $k_{\perp}$, with $Q^2 = n^2 = 0$ and
$\spro{k_{\perp}}{Q} = \spro{k_{\perp}}{n} = 0$ and introduce
\bqa
k_{\mu} &=& z\,Q_{\mu} + k_{\perp\mu} - \frac{k^2_{\perp}}{2\,z}\,
\frac{n_{\mu}}{\spro{Q}{n}},  \nl
p_{i\mu} &=& ( 1 - z)\,Q_{\mu} - k_{\perp\mu} - \frac{k^2_{\perp}}{2\,
( 1 - z)}\,\frac{n_{\mu}}{\spro{Q}{n}},
\eqa
giving
\bq
p^2_i = k^2 = 0, \qquad 2\,\spro{p_i}{k} = - \frac{k^2_{\perp}}{z( 1 -z)}.
\eq
Using the relation
\bq
p_i + k = Q + \ord{k^2_{\perp}},
\eq
we derive
\bq
\sum_{\rm spins}\,\mid \spro{M}{\epsilon}\mid^2 = 2\,Q^2_i e^2\,
\lpar 1 - z\rpar\,\frac{1+(1-z)^2}{k^2_{\perp}}\,\sum_{\rm spins}\,
\mid M_0\lpar p_i \to p_i+k\rpar\mid^2 + \ord{1}.
\eq
The above result shows factorization of the collinear divergence. This 
representation solves many problems, for the leading $k_{\perp}$ 
behavior we must consider only external fermions and we do not need to have
a precise knowledge of the residuals $R^i$. Therefore we do not care about 
including internal $\wb$-bosons emitting photons.

However, the procedure is not gauge invariant, gauge violation occurring at
$\ord{k^2_{\perp}}$, sub-leading w.r.t. leading $\ln k^{\perp}$ corrections.
There are two possibilities. Either photons are allowed only within a cone
(with half-opening $\delta$) surrounding each charged external fermion and we 
identify a leading, gauge-invariant, $\ln(E\delta/m)$ behavior with
sub-leading gauge non-invariant contributions heavily suppressed or we 
integrate over the whole phase space of the photon. For the latter we may
still identify and compute collinear logarithms without having to compute
the exact $\ord{\alpha}$ matrix element but the scale in the logarithm
becomes arbitrary.

Therefore, a rigorous result cannot do without the exact, $\ord{\alpha}$
matrix element and any approximation is not free from ambiguities.
\section{Conclusions\label{concu}}

One of the main ingredients in all calculations aimed to a very accurate
control of high-energy-physics observables is represented by the
re-summation of large QED corrections. This procedure is usually performed
by introducing structure functions. The scale that controls the large
logarithms to be re-summed, as well as the $K$-factor which one
introduces to increase the accuracy of the calculation are based on some
algorithm where one starts from the evolution equation for the structure 
function itself and seek for a solution which factorizes the re-summed
Gribov-Lipatov~\cite{gls} term and which, through an iterative method, matches 
with the finite, second order, result of Barbieri, Mignaco and 
Remiddi~\cite{Barbieri:1972as}. The procedure is to some extent based on
the soft limit.

The are are two classes of problems when one wants to generalize this
algorithm to more complex processes, with many fermions in the final state.
First of all the choice of the scale describing the evolution of the
structure function is ambiguous and it is not even clear that one can
have a realistic description with just one scale in situations where
the dominant contribution to the process is far from the asymptotic
regime. Secondly, an exact fixed order calculation is generally missing for
the process and not only the scale in the dominant logarithms is
ambiguous but also one has no control over the sub-leading logarithms.

From this point of view, all claims that are based on general arguments
as factorization theorems or renormalization group equation are usually
void. The only safe and rigorous argument that one can apply is the 
exact soft-photon re-summation, as given in the classic YFS treatment.

The YFS algorithm has been further developed and modelled for its use in
MonteCarlo programming and we have no pretension to be adding any substantial 
improvement, as clearly stated in the introduction. At the same time
we cannot offer any claim pointing to the complete solution of the
problem of a very precise implementation of QED radiation for processes
more complex than $e^+e^-$-annihilation into fermion-antifermion pairs.
This simple statement should not be confused with a failure of the method.
In \sect{general} we have repeated the classical argument that perturbation
theory and exponentiation can me made consistent with residuals that are
infrared finite. The accuracy at stake is confined in those ingredients that
are missing just because of some technical inadequacy in controlling the
full content of higher orders.

What we have done is an attempt to systematize all arguments and speculations
that have appeared in recent times in the literature. We started with
the well-known YFS re-summation procedure, expressed in the modern
language of dimensional regularization for infrared divergences. Based
on this result we have adopted a slightly different approach where we,
nevertheless, pursue the simple picture in which the whole effect of 
soft-photon emission is described by a real-photon spectral weight function.
However, in our approach we avoid the introduction of a cutoff. 
In this respect we follow an old proposal by Chahine~\cite{Chahine:1978ai}
by introducing an approximation to the exact spectral function which
retains the important properties and incorporates the expected peaking of the
emitted photons along the direction of charged particles. Also in this
case we have completely reformulated the algorithm in modern language.

The result of our investigation allows to write a corrected cross-section
where the kernel for the hard scattering is convoluted with generalized
structure functions where each of them is no longer function of one
scale. Each external, charged, fermion leg brings a factor 
$x^{\scriptscriptstyle{\alpha A - 1}}$ where $\alpha$ is the fine-structure
constant and $A$ is a function which depends on the momenta of the charged
particles.

A preliminar account of these results has been given 
in~\cite{Passarino:2001hz}.

\section{Acknowledgements}

I am grateful to Alessandro Ballestrero for a close collaboration in a early
stage of this project and I thank him and Roberto Pittau for helpful 
discussions and comments on the manuscript.

\section{Appendix A\label{appen}}
The evaluation of the two-particle radiation function is based on Fourier
cosine(sine) transforms of Bessel functions~\cite{bat},
\bqa
g(y) &=& \int_0^{\infty}\,dx\,\cos(xy)\,x^{2\mu-1}\,J_{2\nu}(\alpha x),  \nl
0 \le y \le \alpha, g(y) &=& \frac{2^{2\mu-1}\,\alpha^{-2\mu}\,\egam{\mu+\nu}}
{\egam{1+\nu-\mu}}\,{}_2F_1\lpar \nu+\mu,\mu-\nu;\frac{1}{2};
\frac{y^2}{\alpha^2}\rpar,  \nl
\alpha \le y \le \infty, g(y) &=& \frac{(a/2)^{2\nu}\,y^{-2\nu-2\mu}\,
\egam{2\nu+2\mu}\,\cos(\nu\pi+\mu\pi)}{\egam{2\nu+1}}  \nl
{}&\times& 
{}_2F_1\lpar \nu+\mu,\nu+\mu+\frac{1}{2};2\nu+1;\frac{\alpha^2}{y^2}\rpar,
\eqa
which is valid for
\bq
-\,\Reb\nu < \Reb\mu < \frac{3}{4}.
\eq
\bqa
g(y) &=& \int_0^{\infty}\,dx\,\sin(xy)\,x^{2\mu-1}\,J_{2\nu}(\alpha x),  \nl
0 \le y \le \alpha, g(y) &=& 4^{\mu}\,\alpha^{-2\mu-1}\,y\,\frac{
\egam{\frac{1}{2}+\nu+\mu}}{\egam{\frac{1}{2}+\nu-\mu}}]\,
{}_2F_1\lpar \frac{1}{2}+\nu+\mu,\frac{1}{2}+\mu-\nu;\frac{3}{2};
\frac{y^2}{\alpha^2}\rpar,  \nl
\alpha \le y \le \infty, g(y) &=& \lpar\frac{a}{2}\rpar^{2\nu}\,y^{-2\nu-2\mu}\,
\frac{\egam{2\nu+2\mu}}{\egam{2\nu+1}}\,\sin(\nu\pi+\mu\pi)  \nl
{}&\times&
{}_2F_1\lpar \frac{1}{2}+\nu+\mu,\nu+\mu;2\nu+1;\frac{\alpha^2}{y^2}\rpar,
\eqa
which is valid for
\bq
-\,\Reb\nu - \frac{1}{2} < \Reb\mu < \frac{3}{4}, \qquad  \alpha > 0.
\eq
Another useful integral is~\cite{bat}
\bq
\int_{-1}^{+1}\,dx \exp\lpar i\,zx\rpar\,\lpar 1 - x^2\rpar^{\nu-1/2} =
2^{\nu}\,\pi^{1/2}\,\egam{\nu+\frac{1}{2}}\,z^{-\nu}\,J_{\nu}(z),
\eq
which is valid for
\bq
\Reb\nu > -\,\frac{1}{2}.
\eq
Further, the following integral representation holds~\cite{bat}
\bq
\epsi{s+\alpha} - \epsi{s+\beta} = \intfx{x}\,x^{s-1}\,
\frac{x^{\beta}-x^{\alpha}}{1-x}, \quad \Reb s > - \Reb\alpha,-\Reb\beta.
\eq
We have used two different integral representations for the 
$\Gamma$-function~\cite{bat},
\bq
z^{-s}\,\egam{s} = \int_0^{\infty}\,dx\,x^{s-1}\,e^{-zx}, \qquad
\Reb z > 0, \quad \Reb s > 0,  
\eq
and a second one due to Cauchy and Saalsch\"utz~\cite{ww}
\bq
z^{-s}\,\egam{s} = \int_0^{\infty}\,dx\,x^{s-1}\,\Bigl[
e^{-zx} - \sum_{m=0}^{n}\,\frac{(-zx)^m}{m!}\Bigr],
\eq
which is valid for
\bq
\Reb z > 0, \qquad -(n+1) < \Reb s < -n.
\eq

\end{document}